\documentclass[11pt]{article}
\usepackage[colorlinks]{hyperref}
\usepackage{cite}

\usepackage{amsmath,amsfonts,amsthm,amssymb}
\usepackage{epsfig}  %% TENTATIVA - TEM CLASH WITH PACkAGE GRAPHICS
\usepackage{subfigure}%\usepackage{subcaption}
\def\ii{\'\i}

\def\cao{\c c\~ao}

\def\ii{\'\i}

\def\cao{\c c\~ao}

\def\ftoday{{\sl {Le \number\day \space\ifcase\month 
\or janvier\or f\'evrier\or mars\or avril\or mai
\or juin\or juillet\or ao\^ut\or septembre\or octobre
\or novembre \or d\'ecembre\fi\space \number\year}}}    
\def\ptoday{{\sl {\number\day \space de\space \ifcase\month 
\or janeiro\or fevereiro\or mar{\c c}o\or abril\or maio
\or junho\or julho\or agosto\or setembro\or outubro
\or novembro \or dezembro\fi\space de\space \number\year}}}    
\def\gtoday{{\sl {Den \number\day. \ifcase\month 
\or Januar\or Februar\or M\"arz\or April\or Mai
\or Juni\or Juli\or August\or September\or Oktober
\or November \or Dezember\fi\space \number\year}}}    
\def\today{{\sl {\ifcase\month
\or January\or February\or March\or April\or May
\or June\or July\or August\or September\or October
\or November \or December\fi \space\number\day,\space 
                                            \number\year}}}
\renewcommand{\a}{\alpha}

\newcommand{\g}{\gamma}           
\renewcommand{\d}{\delta}         
\newcommand{\e}{\varepsilon}

\newcommand{\la}{\lambda}        
\newcommand{\m}{\mu}
\newcommand{\n}{\nu}
\newcommand{\om}{\omega}         
\newcommand{\p}{\psi}              
\newcommand{\s}{\sigma}           
\newcommand{\f}{{\phi}}           \newcommand{\F}{{\Phi}}

\newcommand{\CC}{{\cal C}}

\newcommand{\RR}{{\cal R}}
\newcommand{\TT}{{\cal T}}

\newcommand{\es}{\\[3mm]}

\newcommand{\sla}{\raise.15ex\hbox{$/$}\kern -.57em} 
\newcommand{\Sla}{\raise.15ex\hbox{$/$}\kern -.70em}

\newcommand{\lp}{\left(}\newcommand{\rp}{\right)}
\newcommand{\lc}{\left[}\newcommand{\rc}{\right]}
\newcommand{\lac}{\left\{}

\newcommand{\complex}{{\kern .1em {\raise .47ex
\hbox {$\scriptscriptstyle |$}}
    \kern -.4em {\rm C}}}
\newcommand{\real}{{{\rm I} \kern -.19em {\rm R}}}
\newcommand{\rational}{{\kern .1em {\raise .47ex
\hbox{$\scripscriptstyle |$}}
    \kern -.35em {\rm Q}}}
\renewcommand{\natural}{{\vrule height 1.6ex width
.05em depth 0ex \kern -.35em {\rm N}}}

\newcommand{\pa}{\partial}
\newcommand{\pad}[2]{{\dfrac{\partial #1}{\partial #2}}}

\newcommand{\dint}{\displaystyle{\int}}
\newcommand{\eg}{{\em e.g.,\ }}

\newcommand{\ie}{{{\em i.e.},\ }}

\newcommand{\twiddle}{\lower.9ex\rlap{$\kern -.1em\scriptstyle\sim$}}

\newcommand{\ket}[1]{\left| {#1}\right\rangle}
\newcommand{\vev}[1]{\left\langle {#1}\right\rangle}
\newcommand{\equ}[1]{(\ref{#1})}
\newcommand{\eq}{\begin{equation}}
\newcommand{\eqn}[1]{\label{#1}\end{equation}}
\newcommand{\eea}{\end{eqnarray}}
\newcommand{\eqa}{\begin{eqnarray}}
\newcommand{\eqan}[1]{\label{#1}\end{eqnarray}}
\newcommand{\ba}{\begin{array}}
\newcommand{\ea}{\end{array}}
\newcommand{\eqac}{\begin{equation}\begin{array}{rcl}}
\newcommand{\eqacn}[1]{\end{array}\label{#1}\end{equation}}

\makeatletter 
\@addtoreset{equation}{section}  %Reset equation numbering after each section
\makeatother 
 
\newcommand{\bz}{\begin{enumerate}}
\newcommand{\ez}{\end{enumerate}}

\newcommand{\bx}{{\mathbf{x}}}

\newcommand{\yR}{y_{\rm R}}
\newcommand{\yL}{y_{\rm L}}
\newcommand{\VR}{V_{\rm R}}
\newcommand{\VL}{V_{\rm L}}
\title{Quantum Charged Spinning Massless Particles \es
in 2+1 dimensions}

\author{\Large Ivan Morales$^{1}$, 
Bruno Neves$^{1,\, 2}$,
Zui Oporto$^{1,\, 3}$
and 
Olivier Piguet\footnote{Present address: Pra\c ca Graccho Cardoso, 76/504, 45015-180 Aracaju, SE, Brazil}$^{\ ,1}$
}

\begin{document}

\date{}
%\date{October 2019}    %DESMARCAR PARA A DATA APARECER

\maketitle
\vspace{-5mm}

%%%%%%%%%%%%%%%%%%%%%%%%%%%%%%%%%%%%%
\begin{center}
\noindent
$^1$Departamento de F\ii sica, Universidade Federal de 
Vi\c cosa (UFV)
Vi\c cosa, MG, Brazil.\\
$^2$Department of Astrophysics, Cosmology and Fundamental Interactions,
Brasilian Cen\-ter for Research in Physics,
CBPF, {BR-22290-180} Rio de Janeiro, Brazil.\\
$^3$Carrera de F\ii sica, Universidad Mayor de San Andr\'es, La Paz, Bolivia.

%\vspace{0mm}

{\small\tt E-mails:
mblivan@gmail.com, bruno.lqg@gmail.com, \\
azurnasirpal@gmail.com, opiguet@yahoo.com }
\end{center}

%%%%%%%%%%%%%%%%%%%%%%%%%%%%%%%

\vspace{3mm}

%%%%%%%%%%%%%%%%%%%%%%%%%%%%%%%%%%%%%%%%%%%%%%%%%%%%%
\begin{abstract}

 Motivated by the conduction properties of graphene discovered and studied in the last decades, we consider the quantum dynamics of a massless, charged, spin 1/2 relativistic particle in three dimensional space-time, in the presence of an electrostatic field in various configurations such as step or barrier potentials and generalizations of them. The field is taken as   parallel to the $y$ coordinate axis and vanishing outside of a band parallel to the $x$ axis.  The classical theory is reviewed, together with its canonical quantization leading to the Dirac equation for a 2-component spinor.  Stationary solutions are numerically found for each of the field configurations considered, from which we calculate the mean quantum trajectories of the particle and compare them with the corresponding classical trajectories, the latter showing a classical version of the Klein phenomenon. Transmission and reflection probabilities are also  calculated, confirming the Klein phenomenon.

\end{abstract}


%\newpage

%%%%%%%%%%%%%%%%%%%%%%%%%%%%%%%%%%%%%%%%%%%%%%%%
\tableofcontents

%%%%%%%%%%%%%%%%%%%%%%%%%%%%%%%%%%%%%%%%%%%%%%%%%%%%%
%%%%%%%%%%%%%%%%%%%%%%%%%%%%%%%%%%%%%%%%%%%%%%%%%%%%%
\section{Introduction}	

By the end of 1928, the same year that Dirac published its ``quantum
theory of the electron'' paper, Klein~\cite{Klein:1929zz,Calogeracos:1999yp} 
analysed the behaviour of quantum
relativistic electrons in presence of a step barrier potential. His
calculations revealed that, for a sufficiently high enough energy
barrier, electrons can push forward against it and trespassing to
the classical forbidden region by switching the sign of their kinetic
energy, the latter a by-product of the Dirac Hamiltonian having negative
energy states in its spectrum. 
This theoretical prediction is the
 so-called  Klein phenomenon or Klein tunnelling, characterized 
by the absence of the typical exponential
suppression found in  non-relativistic quantum mechanics.

During the analysis of the transmission and reflection coefficients,
Klein apparently also found on a first approach that more electrons
were scattered back by the potential when compared with the number
of incident ones. This implied a reflection coefficient greater than
one and in consequence a negative transmission coefficient. But thanks
to an  insightful observation by Pauli, as Klein himself acknowledged, this
situation was resolved by noticing that inside the barrier potential
the momentum changes its direction, being here opposite to its group velocity.
The modification suffices to restore the physical meaning of the reflection
and transmission coefficients. It has to be mentioned that although
Klein never considered this last result a real paradox -- and in fact
that word does not appear in his paper -- this phenomenon
is many times still known as the ``Klein paradox''.
A very interesting historical 
approach to the Klein gedanken experiment can be seen 
in~\cite{Calogeracos:1999yp, stander:2009, Dragoman_2008, Katsnelson_2006}. 

It is by now widely understood~\cite{Calogeracos:1999yp}  
that this phenomenon
is a pure relativistic effect. 
 In the regions where 
the total energy $E$ is smaller than the critical value 
$V(\mathbf{x})-m c^2$,
\ie where the kinetic energy is negative,
the particle still propagates, with an oscillating wave function. 
In a solid state context,  this can be interpreted as a particle (an electron) propagating in the valence band, whereas in the region where the potential is such that the kinetic energy is positive,  $E>$ 
$V(\mathbf{x})+m c^2$, one has a particle propagating in 
the conduction band. We shall refer to such situations as the
propagation of a VB, respectively CB, particle 
(see figure 2 in~\cite{Calogeracos:1999yp}).
 Thus  a relativistic particle 
 may go through a potential 
step or barrier  without  exponential damping irrespective 
of the energy value, provided $|E-V|>mc^2$,  
in contrast to the non-relativistic case. 
The classical counterpart of this effect is that the potential is 
repulsive or attractive  depending on the sign of the kinetic energy (see Appendix \ref{app-electrostatic field}).

By the time Klein published his work, it was conjectured that the
``paradoxical'' result mentioned above was caused by the abrupt discontinuity
of the step potential. Could a smoother potential get rid of this unintuitive
result? In this respect, it was shown by Sauter~\cite{Sauter_1931} , 
that the
conjecture was partially correct: for weak smooth fields the reflection
and transmission coefficients behaves as expected, but for strong
fields Klein phenomenon shows up despite of the potential being continuous.

Later, Hund~\cite{Hund_1941}  
reconsidered the analysis from the quite different
perspective of multiparticle theory in quantum field theory. Although
his calculations were limited to the Klein-Gordon equation, it was
clear by the time that the potential barrier was spontaneously producing
pairs of charged 
 particles/antiparticles. The analysis for the Dirac
field was successfully accomplished  fifty years later by  
Nikishov~\cite{Nikishov:1969tt,Nikishov:1969ttENG,Nikishov:1970br}).
 Since then, 
the Klein phenomenon has been adopted
by some authors as a good and pedagogical starting point to justify
the introduction of the multi-particle picture of the Dirac equation
against the single particle stand point. 
A closer examination of the Klein phenomenon shows
that this is not necessarily true, since Klein tunnelling can occur
even when the potential is not high enough to produce 
pairs~\cite{Calogeracos:1999yp}. It has also been suggested to trace Klein tunnelling
back to the classical relativistic 
theory~\cite{Hansen_1981, Beenakker:2008zz}  
where massless 
charged particles possess a very special dynamics in the presence
of external electromagnetic fields. Let us mention in particular
a result implicit
in~\cite{Morales:2017rlk, Hansen_1981, Beenakker:2008zz} 
 that for a spinless particle
travelling with velocity $\mathbf{v}=(0,v_{0y}=c,0)$ right towards a
 constant electric field $(0,E,0)$ barrier, no matter its intensity,
the particle experience no back scattering at all (see equation (24) 
of~\cite{Morales:2017rlk}).
 The explanation of this unexpected
result is clear if we remember that the massless and spinless particle
already moves at the speed of light; a reflection would imply a turning
point at which the velocity would be zero, which is not admissible.
Therefore there is no such classically forbidden region for this particular
case.
The reader can refer to~\cite{Beenakker:2008zz} 
for further discussions on this point.

It has to be noted that Klein phenomenon (along with the
\textit{Zit\-ter\-be\-we\-gung}) remains as a theoretical 
prediction with no experimental
evidence so far  in high energy physics. This fact seems to have contributed to
keep alive the debate almost a century after Klein's 
work\footnote{An inspection of 
the cross reference database  on the inspirehep.net
web site reveals a
significant increasing of citations to the original Klein article
since 2010 (http://inspirehep.net/record/48390/citations).}. 
A more refreshing
debate about the Klein phenomenon has been gaining attention in recent
years, this time in the context of condensed matter physics, more
specifically the physics of graphene 
(see~\cite{CastroNeto:2009zz} for a review). Indeed,
the band structure of the  monolayer of carbon known as 
graphene has been dem\-ons\-trat\-ed to be a suitable testing ground for
quantum  electrodynamics  phenomena. Its effective 
charge carriers obey a relativistic linear energy-momentum 
dispersion relation  of the form $E=v_{\rm F}|\mathbf{p}|$,
the Fermi critical velocity $v_{\rm F}$ $\approx 1000$ km/s
playing the role of the 
``light  velocity'' $c$. Thus, graphene 
stands out as an exceptional material for testing such 
a special quantum behaviour. In particular, Klein tunnelling has being
experimentally confirmed ten years ago~\cite{stander:2009,Young}.
More generally, a number of devices for 
applications have been proposed in order to investigate 
this phenomenon~\cite{stander:2009, Cheianov:2006aa}.

The significant advances in the physics of 
graphene, first on Klein tunnelling~\cite{Tudorovskiy,Reijnders1,Kleptsyn}, 
then on the quantum Hall effect~\cite{ Zhang:2005zz, Novoselov:2005kj},
on the scattering properties and 
``electronic optics''~\cite{Afranio,Logemann,Reijnders2,
Reijnders3,Reijnders4} 
and also some recent results on the 
so-called topological semi-metals, \textit{e.g.} Weyl 
semi-metals~\cite{Hosur2013857,Lu_2017}, 
are promising for novel applications
for fundamental physics as well as for probing for 
new theoretical and experimental electronic, optical 
and mechanical 
properties.  
%In particular,  a back door has
%opened to test, for the first time, the Klein phenomenon
%among with other interesting quantum field theory analogue phenomena. 
On the other
hand, the novelties of graphene from the theoretical point of view
has not been unnoticed for the quantum gravity 
community~\cite{Guevara:2016rbl, Alvarez:2015bva, Capozziello:2018mqy, Iorio:2013ifa, Iorio:2014nda, Mesaros:2009az, Sepehri:2017dky, Iorio:2018agc}, 
 once again offering a unique opportunity to develop
future experiments to guide theory.

Within the theoretical framework of condensed matter physics Klein
tunnelling can be interpreted as the interband tunnelling (i.e. the
transition of an electron from the conduction band to the valence
band). The interband transition is possible because the presence of
the step barrier modifies the dispersion relation to the 
 right of the
step (See~\cite{Calogeracos:1999yp,Allain:2011wk} and Fig. \ref{CB-VB}). 
The electrons in the conduction
band are analogous of the ordinary electrons with positive energies
in relativistic quantum mechanics, with its velocity and momentum
pointing in the same direction. The electrons in the valence band
are analogous to negative energy electrons; in this case the velocity
and momentum points to opposite directions. A similar description
applies to holes (the absence of an electron in any of the bands).
Holes in the conduction band are the equivalent to positive energy
positrons, with the velocity and momentum pointing to the same direction
(the difference with electrons would be in the direction of the current
because of the sign of the charge) whereas holes in the valence band
corresponds to negative energy positrons. This is a framework compatible
with the single particle picture of the Dirac equation. It is clear
that the problem of charge conservation has no place in this case.
What is most remarkable is that this description already contains
the clue ingredient pointed out almost 100 years ago by Klein and
Pauli, \ie  electrons inside the barrier has negative kinetic energies,
and its momentum experiences a change of sign with respect to the group
velocity.

The existence of such peculiar properties justifies 
further theoretical in\-ves\-tig\-at\-ions through a 
formalism based on fundamental principles, namely, 
from a Lagrangian point of view and
following the Dirac-Bergmann prescription to 
implement the canonical quantization. A complete 
quantum analysis for a spinning, charged and massless 
particle in $(2 + 1)$ dimensions  is still missing, 
although many results may be found in the literature.

For effectively massless fermions such as in graphene, 
there is no energy gap, thus the Klein 
tunnelling, occurs as soon as the energy $E$ is lower than the potential
%inminently 
for direct
incident electrons. For oblique incident ones  the component of 
the momentum parallel to the potential barrier emulates, in some respects,
an effective non zero mass, and thus a non zero gap separating the
negative and positive kinetic energy states reappears, as our 
calculations will confirm  (see Fig. \ref{disp-rel}). 

We must stress that, motivated by the dynamics of electrons 
in materials such as graphene, we deal here with a spin 1/2 massless particle
obeying a Dirac equation, which may be obtained by a quantization 
procedure from a supersymmetric classical theory involving 
Grassmann variables whose quantum operator version is 
represented by  Dirac matrices. 
This must be contrasted with the so-called anyon 
theories~\cite{Hanson-Regge,Chaichian-etall,Ghosh}\footnote{We thank 
Dr. Subir Gosh for appointing these references to us.} 
--  also in 3D space-time -- where the classical theory 
has only ordinary variables and the quantum wave equation 
is not of the Dirac form.
There, spin can take arbitrary values. Moreover, at least 
in the  
references~\cite{Hanson-Regge,Chaichian-etall,Ghosh}, the mass 
is different from zero.

 The aim of the present paper is to discuss the quantum dynamics of 
a relativistic massless charged fermions in (2+1) dimensional space-time,
in the context of Dirac 
wave mechanics. We shall follow a strictly quantum-mechanical 
approach~\cite{Klein:1929zz, Thaller:1992,Calogeracos:1999yp,Dragoman_2008,Katsnelson_2006,Sauter_1931,Beenakker:2008zz,Allain:2011wk,Dombey:1999id,Katsnelson:2012,Tudorovskiy,Reijnders1},
 and not the full quantum field theoretic 
one~\cite{Hund_1941,Nikishov:1969tt,Nikishov:1969ttENG,Nikishov:1970br,Hansen_1981,Calogeracos2,Krekora,OswaldoEmerson}
which would take into account 
processes such as pair production and the possibility of
 e$^-$e$^-$ bound states~\cite{OswaldoEmerson}.

We begin in Section 2 with
the full canonical quantization of the theory, whose 
classical aspects have been  studied in a 
previous work~\cite{Morales:2017rlk}. Therefore, we start with the 
analysis of the constraints for the spinning charged 
particle, taken  as massive. We follow  then
the Dirac-Bergmann 
prescription for the quantization in the massless 
case. 

In Section $3$ we start the study of the quantum  massless particle 
coupled with an external electrostatic potential. 
A technical analysis of the boundary conditions is 
described in order to make basic points more 
transparent and we present the results for various types
of potentials, square or more general ones. 
When necessary, comparisons with the 
literature~\cite{CastroNeto:2009zz, stander:2009, Dragoman_2008, Katsnelson_2006, Calogeracos:1999yp, Dombey:1999id, Thaller:1992, Katsnelson:2012} 
 -- which mainly deals 
with square potentials -- will be made.
 In each case we calculate the reflection and transmission 
 probabilities, and we  compare the quantum mean trajectory of the 
  particle 
with its classical counterpart, as a check of the correspondence
principle. We observe the quantum effect 
\textit{Zitterbewegung}, \ie a jittery motion of the mean trajectory 
 due to the 
superposition of the right and left moving waves -- when both are present.
 Two appendices are dedicated to the definition of conventions and to 
the classical equations of motion in a special case of interest.

Computations in concrete cases are done with the help of the 
software Mathematica~\cite{Wolfram}.  Interested readers may 
download (and use) the computer program from the arXiv site of 
this paper (link: https://arxiv.org/src/1910.03059v2/anc/Trajectories.nb 
and save it as a file: Trajectories.nb).

%%%%%%%%%%%%%%%%%%%%%%%%%%%%%%%%%%%%%%%%%%%%%%%%%%%%%%%%%%%%%%%%%%%%
\section{Canonical analysis and quantization}
\subsection{Action and classical equations of motion}
%============================
The classical motion of a relativistic spinning particle of 
mass\footnote{We will be most interested in the massless particle, 
but in this section we consider the massive case for the sake of 
generalization.}  $m$ and electric charge $q$  in the presence
 of an external electromagnetic field $A_\mu(x)$ reads, in covariant form:
\eq
S=\int_{\mathcal{C}}d\la \, L(X(\la),\dot X(\la)),
\eqn{action4}
where $\CC$ is a path in dimension 3 spacetime\footnote{The spacetime index 
$\m$ takes the values 0, 1, 2, the metric is 
$\eta_{\m\n}$ = diag$(1,-1,-1)$. We use natural units with $c=\hbar=1$ ($c$ would be the Fermi velocity in applications such as graphene physics).}
 parametrized by $\la$. $X$ represents the 
generalized coordinates $x^\m$, $e$, $\p^\m$, $\chi$, $\p_5$, 
the last three ones being odd (anticommuting) Grassmann numbers
which describe the classical spin degree of freedom. 
Dot above the variables denotes derivatives with respect to $\lambda$.
The Lagrangian  is given 
by~\cite{Balachandran:1976ya, Brink:1976sz, Fainberg:1987jr,Morales:2017rlk} 
\eq\ba{l}
	L(X,\dot X)=-\dfrac{1}{2}\left(\dfrac{\dot{x}^{\mu}}{e}
	\left(\dot{x}_{\mu}-i\chi\psi_{\mu}\right)
	- i\psi^{\mu}\dot{\psi}_{\mu}\right)
		-\dfrac{1}{2}\left(e m^{2}
		+ i\left(\psi_{5}\dot{\psi}_{5}+m\chi\psi_{5}\right)\right) \\[3 mm]
		\hspace{30pt}-\left( q A_{\mu}(x)\dot{x}^{\mu}
		+ \dfrac{iq}{2}e\psi^{\mu}F_{\mu\nu}(x)\psi^{\nu}\right),
\ea\eqn{lagrangian}
$A_\m$ being the electromagnetic 3-potential and $F_{\m\n}$ = 
$\partial_{\mu}A_{\nu}-\partial_{\nu}A_{\mu}$ the 
electromagnetic tensor field.
The odd Grassmann variable $\p_5$ may be ommitted in the massless 
case $m=0$.

The action is invariant, up to boundary terms, under two gauge symmetries. 
The first one is its invariance under the $\la$--reparametrizations:
\eq\ba{l}
\delta^{\rm R}_{\varepsilon} x^{\mu}=\varepsilon \dot{x}^{\mu},
\quad \delta^{\rm R}_{\varepsilon} \psi^{\mu}= \varepsilon \dot{\psi}^{\mu},
\quad\delta^{\rm R}_{ \varepsilon} \psi_{5}= \varepsilon\dot{\psi}_{5},\es
\delta^{\rm R}_{\varepsilon} e = \dot{\varepsilon}e+\varepsilon\dot{e} ,
\quad\delta^{\rm R}_{\varepsilon} \chi = \dot{\varepsilon}\chi +  \varepsilon \dot{\chi},\ea\eqn{reparam-transf}
where $\e(\la)$ is an infinitesimal parameter. 
The second gauge invariance is a supersymmetry:
\eq\ba{l}
\delta^{\rm S}_{\alpha} x^{\mu}=i\alpha \psi^{\mu},
\quad\delta^{\rm S}_{\alpha} \psi^{\mu}
 = \alpha\dfrac{1}{e}\lp\dot{x}^{\mu}-\dfrac{i}{2} \chi\psi^{\mu}\rp,
\quad\delta^{\rm S}_{\alpha} \psi_{5}
   =m\alpha+ \dfrac{i}{m e} \alpha \psi_{5}\left(\dot{\psi}_{5}-\dfrac{m\chi}{2}\right), \es
\delta^{\rm S}_{\alpha} e = i \alpha \chi ,
\quad\delta^{\rm S}_{\alpha} \chi=2\dot{\alpha},
\ea\eqn{susy-transf}
where $\a(\la)$ is an odd infinitesimal parameter.
The electromagnetic potential and field accordingly transform as
\[\ba{ll}
 \delta^{\rm R}_{\varepsilon}A_{\mu}=\varepsilon \dot{A}_{\mu} ,
&\quad \delta^{\rm R}_{\varepsilon}F_{\mu\nu}=\varepsilon \dot{F}_{\mu\nu}\es
\delta^{\rm S}_{\alpha}A_{\mu}=i\alpha \partial_{\rho}A_{\mu}\psi^{\rho},
&\quad \delta^{\rm S}_{\alpha}F_{\mu\nu}=i\alpha \partial_{\rho}F_{\mu\nu}\psi^{\rho}
\ea\]
A superalgebra structure is evidenced by the commutation rules
\[
\lc \d^{\rm R}_{\e_1},\d^{\rm R}_{\e_2}\rc 
   =  \d^{\rm R}_{{\e_2}\dot{\e_1}-{\e_1}\dot{\e_2}},
\quad \lc\d^{\rm R}_\varepsilon,\d^{\rm S}_\a\rc  
   = \d^{\rm S}_{-\varepsilon\dot\a},
\quad \lc\d^{\rm S}_{\a_1},\d^{\rm S}_{\a_2}\rc 
=  \d^{\rm R}_{\tilde\varepsilon}+ \d^{\rm S}_{\tilde\a},
\]
where $\tilde\varepsilon=2i{\a_2}{\a_1}/e$ and $\tilde\a=-i{\a_2}{\a_1}\chi/e$. Note that the
structure "constant" in the last commutator depends on the variable $\chi$.

The equations of motion obtained by the variation of the action \equ{action4} read
\eq\ba{l}
\dfrac{\rm d}{d\lambda}\left(\dfrac{\dot{x}_{\mu}}{e}-i\dfrac{\chi\psi_{\mu}}{2e}\right)-q \left(F_{\mu\nu}\dot{x}^{\nu}+ \dfrac{i}{2}e\psi^{\rho}\partial_{\mu}F_{\rho\sigma}\psi^{\sigma}\right)=0,\es
\dfrac{1}{2}\left(\dfrac{\dot{x}_{\mu}\dot{x}^\mu}{e^2}
-i\dfrac{\chi\dot{x}_{\mu}\psi^{\mu}}{e^2}+ i q \psi^{\mu}F_{\mu\nu}\psi^{\nu}\right)-\dfrac{m^2}{2}=0,\es
i\left( \dot{\psi}_{\mu}-\dfrac{\dot{x}_{\mu}\chi}{2e}- q e  F_{\mu\nu}\psi^{\nu}\right)=0,\es
- i\left(\dot{\psi}_{5}-\dfrac{m}{2}\chi\right)=0,\es
\dfrac{i}{2}\left(\dfrac{\dot{x}^{\mu}\psi_{\mu}}{e}- m\psi_{5}\right)=0.
\ea\eqn{eom}
Note that we could choose  $\chi=0$  as a gauge fixing condition, 
which leaves a residual supersymmetry \equ{susy-transf} with 
a constant parameter $\a$, which
in turn could be fixed, in the massive case, by the condition $\p_5=0$. 
This can easily be checked by examining the supersymmetry 
transformations \equ{susy-transf}
and observing that the fourth of the equations \equ{eom} implies 
$\dot\p_5=0$ if $\chi=0$.

 We display  in Appendix 
\ref{app-electrostatic field} the field equations for the  massless 
charged spinning particle in the presence of an electrostatic 
field depending only on the $y$ coordinate, which is the case 
of interest in the application part of the paper.

%%%%%%%%%%%%%%%%%%%%%%%%%%%%%%%%%%%%%%%%%%%%%%%%%%%%%%%%%%%%%

%%%%%%%%%%%%%
%=======================================
\subsection{Canonical analysis}
%=======================================

As a preparation for the quantization of the theory we perform a canonical
analysis following Dirac's algorithm  
for systems with constraints~\cite{Dirac, Henneaux:1992ig, Sundermeyer:1982gv, Gitman:1990qh}.
  We keep a non-zero mass in the present section. 

%%%%%%%%%%%%%%%%%%%%%%%%%%%%%%%%%%%%%%
\subsubsection{Analysis of the constraints}
%%%%%%%%%%%%%%%%%%%%%%%%%%%%%%%%%%%%%%%

The conjugate momenta are  read out from the Lagrangian \equ{lagrangian}:
\eq\ba{c}
p_{\mu}=\dfrac{\pa L}{\pa\dot{x}^{\mu}}
   =-\dfrac{\dot{x}_{\mu}}{e}-q A_{\mu}+\dfrac{i \chi \psi_{\mu}}{2 e},\qquad
p_e=\dfrac{\pa L}{\pa\dot{e}^{\mu}}=0,\\[4mm]
\mathbb{P}_{\mu}=\dfrac{\partial L}{\partial \dot{\psi}^{\mu}}
   =- \dfrac{i \psi_{\mu}}{2},\qquad
p_{\chi}=\dfrac{\partial L}{\partial \dot{\chi}}=0,\qquad
p_{5}=\dfrac{\partial L}{\partial \dot{\psi}^{\mu}}= \dfrac{i \psi_{5}}{2}.
\ea\eqn{conj-mom}
Four of these equations relate momenta and generalized coordinates, 
which means that we have four primary constraints:
\eq\ba{c}
\f_e= p_e\approx0  ,\quad
\f_\chi= p_{\chi} \approx 0,\quad
\f_\m=\mathbb{P}_{\mu}+ \dfrac{i \psi_{\mu}}{2}\approx 0  ,\quad
\f_5=p_{5}- \dfrac{i \psi_{5}}{2}\approx 0,
\ea\eqn{prim-constr}
where the symbol $\approx$ means a ``weak equality'', \ie an equality
which will be turned effective only after all the Poisson bracket algebra 
manipulations are done.

The basic non-vanishing Poisson brackets between the generalized coordinates 
and their conjugate momenta are given by:
\eq\ba{c}
\lbrace{ x^\mu , p_\nu \rbrace} =\delta^\mu_\nu,\quad
 \lbrace{ e , p_e \rbrace} =1,\es
\lbrace{ \psi^\mu , \mathbb{P}_\nu \rbrace} =-\delta^\mu_\nu, \quad
 \lbrace{ \chi , p_\chi \rbrace} =-1,\quad
\lbrace{ \psi_5 , p_5 \rbrace} = -1.
\ea\eqn{nzbp}
These brackets are ``graduated'', \ie they are symmetric if both arguments are Grassmann odd, and antisymmetric otherwise.

Through a Legendre transformation we obtain the canonical Hamiltonian
\eq
H_{\rm C} = -\dfrac{e}{2}\phi_{\rm KG}+\dfrac{i \chi}{2}\phi_{\rm D}
\eqn{can-Hamiltonian}
where
\eq
\phi_{\rm KG}=\Pi^{\mu}\Pi_{\mu}-m^{2}- i q \psi^{\mu}F_{\mu\nu}\psi^{\nu},
\qquad \phi_{\rm D}=\psi_{\mu}\Pi^{\mu}+ m\psi_{5},
\eqn{second-constr}
with
\[
\Pi_{\mu}=
 p_{\mu}+q A_{\mu}=-\dfrac{\dot{x}_{\mu}}{e}+\dfrac{i \chi \psi_{\mu}}{2 e}.
 \]
Both terms of the canonical Hamiltonian turn out to be 
secondary constraints assuring the stability of the primary constraints
$\f_e$ and $\f_\chi$:
\eq 
\f_{\rm KG}\approx 0,\qquad \f_{\rm D}\approx 0.
\eqn{sec-constr}
One first   notices  that the last two constraints in \equ{prim-constr} 
have Poisson brackets which do not weakly vanish:
\eq 
\{\f_\m,\f_\n\}=-i\eta_{\m\n},\quad \{\f_5,\f_5\}=i,\quad
\{\f_\m,\f_5\}=0,
\eqn{2classPB}
which means that they are second class. 
These Poisson brackets
form a non-singular $4\times 4$ matrix $C_{AB}$, with
$A=\m,5$ and $B=\n,5$:
\eq 
C_{AB}=\lp \ba{cc}-i\eta_{\m\n}&0 \\ 0& i\ea\rp.
\eqn{matrix-C}
Elimination of the second class constraints $\f_A$ is performed by 
introducing the Dirac brackets
\eq 
\{U,V\}_{\rm D}=\{U,V\}
 - \sum_{A,B}\{U,\f_A\}\lp C^{-1}\rp^{AB}\{\f_B,V\},
\eqn{Dirac-brackets}
for any  phase space functions  $U$ and $V$. 
The Dirac brackets of the second class constraints with any phase space function
being strongly vanishing,
this allows one to solve them right now:
\eq 
\mathbb{P}_\m=-\dfrac{i}{2}\p_\m,\quad p_5=\dfrac{i}{2}\p_5.
\eqn{sol-2nd-class}
Phase space is thus reduced, its coordinates being now
$x^\m,\,p_\m,\,\p_\m,\,e,\, p_e,\,\chi,\, p_\chi$ and $\p_5$.
The fundamental non-zero Dirac brackets for our system are then given by:
\eq\begin{array}{c}
	\lbrace{ x^\mu , p_\nu \rbrace}_{\rm D}\hspace{0pt}=\delta^\mu_\nu, \quad
 \lbrace{ e , p_e \rbrace}_{\rm D} \hspace{9pt}=1,\es
\lbrace{ \psi_\mu , \psi_\nu \rbrace}_{\rm D}\hspace{0pt} =- i\eta_{\mu\nu},\quad
\lbrace{ \chi , p_\chi \rbrace}_{\rm D}=-1,\quad
 \lbrace{ \psi_5 , \psi_5 \rbrace}_{\rm D} = i.
\end{array}\eqn{Dbracket1}

Finally one checks that the Poisson brackets between the 
 constraints \equ{sec-constr},   
the first two of \equ{prim-constr} and the 
Hamiltonian \equ{can-Hamiltonian}  are weakly zero, \ie 
they are either zero or a linear combination of constraints. 
This means that they are ``first class''.   In particular
they  are left
stable during their evolution with respect to the  
``time'' $\la$, generated by their brackets with the Hamiltonian. 
Thus, no new constraint occurs: the present set of constraints 
is  complete.

%%%%%%%%%%%%%%%%%%%%%%%%%%%%%%%%%%%%%%
\subsubsection{Gauge invariances}
%%%%%%%%%%%%%%%%%%%%%%%%%%%%%%%%%%%%%%%

Each first class constraint $\f_A$ 
($A= {\rm KG},\ {\rm D},\ e,\ \chi$)
generates an invariance under a gauge transformation which, infinitesimally, 
takes the form
\eq 
\d_{A} U=\e\{U,\f_A\}_{\rm D},
 \eqn{canonical-g-tr}
for any function $U$ of the reduced phase space, with $\e$ an infinitesimal parameter.
We see from \equ{prim-constr}, that $\f_e$ and $\f_\chi$ generate 
arbitrary translations of the coordinates $e$ and $\chi$, respectively. 
This implies that we can gauge fix each of them to an 
arbitrary function. We can then read the coefficients 
$e$ and $\chi$ in the canonical Hamiltonian \equ{can-Hamiltonian} as 
Lagrange multipliers for the constraints $\f_{\rm KG}$ and $\f_{\rm D}$, 
and forget the constraints $p_e$ and $p_\chi$.

We are left with a reduced phase space of coordinates
$x^\m,\,p_\m,\,\p_\m$ and $\p_5$
and two gauge invariances generated by  
$\f_{\rm KG}$ and $\f_{\rm D}$, obeying the Dirac bracket algebra
\eq
\{\f_{\rm D},\f_{\rm D}\}_{\rm D}=-i\f_{\rm KG},
\eqn{D-bracket-alg}
the other brackets being vanishing. They generate the gauge 
transformations, 	according to \equ{canonical-g-tr}:
\eq\ba{lll}
\d_{\rm KG}\,x^\m
 = -\dfrac{2\e}{e}\dot{x}^\m,
\quad& \d_{\rm KG}\,\p^\m = -2\e qF^\m{}_\n\p^\n, &\quad \d_{\rm KG}\,\p_5=0 \es
\d_{\rm D}\,x^\m = \a\p^\m,\quad& \d_{\rm D}\,\p^\m =-i\a\dot{x}^\m , 
&\quad \d_{\rm D}\,\p_5 = i\a m
\ea\eqn{g-transf-KG-D}

%===========================================
\subsection{Quantization}
%===========================================

From now on we consider the massless case $m=0$ and 
therefore one can take $\p_5=0$.
 
 To convert the classical theory to its quantum version, we proceed according 
 to the Dirac scheme~\cite{Dirac}, promoting the  classical expressions $A$  
 to operators $\hat A$, and then impose  (anti-)commutation relations  
 on these operators. These (anti-)commutation relations may be viewed as the outcome of the substitution of the Dirac bracket 
 $\{A,B\}_{\rm D}$ defined in the preceding section by the graded commutator 
 $(i \hbar)^{-1}[\hat{A}, \hat{B}]$. In particular, from (\ref{Dbracket1}), we find the basic (anti-)commutation relations  between the  canonical variables:
 \begin{equation}
	\lbrack  \hat{x}^\nu, \hat{p}_\mu \rbrack =i\hbar \delta_\mu^\nu ,\quad
 \lbrack \hat{\psi}_\mu, \hat{\psi}_\nu \rbrack  \hspace{2pt}= \hbar \eta_{\mu\nu},
 %\\[3 mm]
%{\color{blue}\lbrack \hat{\psi}_5, \hat{\psi}_5 \rbrack_+} \hspace{4pt}=- \hbar.\\[3 mm]
	\label{commutationR1}
\end{equation}
In a wave mechanics representation,
the state is described by a 
2-components\footnote{ The present quantization procedure 
with 2-components spinors holds 
for the massless case. 
In the massive case there would be 4 Dirac matrices  obeying a 
Clifford algebra, corresponding to the 
4 odd variables $\p_\m,\ \p_5$; the dimension of the representation 
of this algebra would thus been at least 4.}
spinor $\Psi(x)$, and the basic operators are defined as
\[
\hat{x}^\mu=x^\mu,\quad
\hat{p}_\mu=-i\partial_\mu,\quad
\hat{\psi}^\mu=\gamma^\mu/ \sqrt{2},
\]
where the $\g^\m$ are the 2-dimensional Dirac matrices, with
$\g^\m\g^\n+\g^\n\g^\m=2\eta^{\m\n}$.

With this prescription, the first class constraint $\phi_D$ defined in \equ{second-constr} 
with $m=0$, yields the Dirac wave equation:
\begin{equation}
\label{Deq1}
\hat{\phi}_D\Psi(x)=\gamma^\mu\left(-i\partial_\mu+qA_\mu\right)\Psi(x)=0.
\end{equation}

%%%%%%%%%%%%%%%%%%%%%%%%%%%%%%%%%%
\section{The quantum relativistic massless particle in an electrostatic 
potential}\label{electrostatic potential}
\subsection{General setting}

We consider the Dirac equation \equ{Deq1} for a massless 
spin 1/2 particle of unit charge $q=1$ in a static electric 
field\footnote{See Appendix \ref{not-conv} for our notations 
and conventions.}
$\mathbf{E}=-\nabla V$, $V(\mathbf{x})=A_0(\mathbf{x})$ being the electric potential.
It reads
\eq
\lp\g^0\lp i\pad{}{t}-V(\bx)\rp +i\,\g^i\pad{}{x^i} \rp \Psi(t,\bx) = 0.
\eqn{Dirac-eq}
The Dirac Hamiltonian operator is given by
\eq
H_{\rm D}=-i\a^i\pad{}{x^i}+V(\bx).
\eqn{Hamiltonian}
The matrices $\g^\m$ and $\a^i$ are given in Appendix \ref{not-conv}.

The density and flux of probabilities are given by the components of the 
3-current
\eq\ba{l}
J^\m(x)=\bar{\Psi}(x)\g^\m\Psi(x),\es
\rho(t,\bx)=J^0(t,\bx)=\Psi^\dagger(t,\bx)\Psi(t,\bx),
\quad J^i(t,\bx)=\Psi^\dagger(t,\bx)\a^i\Psi(t,\bx),\quad i=1,2.
\ea\eqn{3-current}
In order to compare the theory with the classical one we will need to 
compute the mean position and velocity of the particle, which we 
will denote by $\mathbf{x}_{\rm q}$ = $(x_{\rm q},y_{\rm q})$ and 
$\mathbf{v}_{\rm q}$ =  $(v_{\rm q}^x,v_{\rm q}^y)$, 
respectively. In a general
state described by a (normalizable) vector $\ket{\Psi}(t)$, 
the mean velocity is given by 
\eq
v_{\rm q}^i(t) = \dfrac{1}{||\Psi||^2}\vev{\Psi|\a^i|\Psi}(t),\quad i=1,2,
\eqn{mean-velocity-general}
and the mean position by integrating each side of the latter equation.
This expression amounts to integrate the fluxes $J^i$ and divide 
by the integral of the density $\rho$  given in \equ{3-current}.

However, when dealing with (non-normalizable) 
 stationary scattering states, we will
use the following alternative. We first define ``local mean velocities''
\eq 
v_{\rm q}^i(\bx)= J^i(\bx)/\rho(\bx),
\eqn{local-velocity}
and then find the mean position $x_{\rm q}^i(t)$ by integrating 
 with suitable boundary conditions  the differential equations
\eq
{\dot x_{\rm q}}^i(t) = v_{\rm q}^i(\mathbf{x_{\rm q}}(t)),
\eqn{q-eqs}
where a dot means time differentiation.

In this section, we restrict ourselves to the special case of 
a potential depending only on the $y$-coordinate: $V=V(y)$,
the electric field $\mathbf{E}(y)=(0,\,-V'(y))$ being 
parallel to the $y$-axis. Moreover, 
the electric field is assumed to vanish outside of an interval 
$\yL\le y\le y_{\rm R}$. More precisely, the potential  obeys the conditions
\eq
V(y)=\left\lbrace\ba{ll}
\VL,&\quad y\le \yL,\\
\VR,&\quad y\ge \yR,
\ea\right.
\eqn{compact-V}
where $\VL$ and $\VR$ are constant. An illustration is provided by Fig. \ref{potgeneric}.

We take the 
two-component Dirac spinor to be stationary and, due to the 
$x$-in\-dep\-en\-den\-ce of the potential, to be an eigenvector of the 
$x$-momentum component, 
with eigenvalue $k_x$:
\eq
\Psi(t,x,y)=\lp\ba{l}\F_1(t,x,y)\\ \F_2(t,x,y)\ea\rp
=e^{-i\om t+i k_x x}\lp\ba{l}f(y)\\ g(y)\ea\rp,
\eqn{Dirac spinor}
where $\om$ is an eigenvalue of the Dirac Hamiltonian operator
\equ{Hamiltonian}.
The functions $f(y)$ and $g(y)$ then obey the equations
\eq\ba{l}
i(V(y) - \om) f(y) - k_x g(y) + g'(y)=0,\es
 i(V(y) - \om) g(y) + k_x f(y) + f'(y)=0,
\ea\eqn{component-Dirac-eq}
and the probability density and flux read
\eq\ba{c} 
\rho(y) = f^*(y)f(y)+g^*(y)g(y)\es
J^x(y)= i\lp  f^*(y)g(y)-g^*(y)f(y) \rp,\quad
J^y(y)=   f^*(y)g(y)+g^*(y)f(y), 
\ea\eqn{density-and-current}
where $^*$ means complex conjugation. In the present case, the local 
velocity  \equ{local-velocity} entering in the differential 
equations \equ{q-eqs} depends only on the coordinate $y_q(t)$.

%%%%%%%%%%%%%%%%%%%%%%%%%%%%%%%%%%
\subsubsection{Asymptotic states}
\label{asymptotic states}

Outside of the interval 
$\yL<y<y_{\rm R}$, the wave function obeys the Dirac equations 
 \equ{component-Dirac-eq} with a constant electrostatic potential, 
 denoted by $V$ in the present subsection.
 Its solutions are progressive  
 waves $(f^V_{\pm},\,g^V_{\pm})$, where the suffix $\pm$ means
  right or left mode, respectively. Up to an overall factor:
\eq
f^V_{\pm}(y) = e^{\pm ik_y y}, \quad
g^V_{\pm}(y) = \dfrac{\pm k_y-ik_x}{\om-V} e^{\pm ik_y y},
\eqn{right-free-waves}
%%%%%%%%%%%%%%%%%%%%%%%%%%%%%%%%%%%%%%%%%%%%%%%
\begin{figure}[htb]
\centering
\subfigure[The figure shows the dispersion relation for
massless particles in 2+1 dimensions (the double conic surface). 
For $k_x$ fixed, the particle 
is constrained to move along the hyperbola (dashed curve) 
that lies in the plane $k_x$ = const.; in this case the effective 
dispersion relation is $\om(k_y) = \pm \sqrt{k_y^2 + k_x^2}$, which is 
equivalent to the dispersion relation of a particle of mass $|k_x|$ 
in natural units.]{
\includegraphics[scale=0.25]{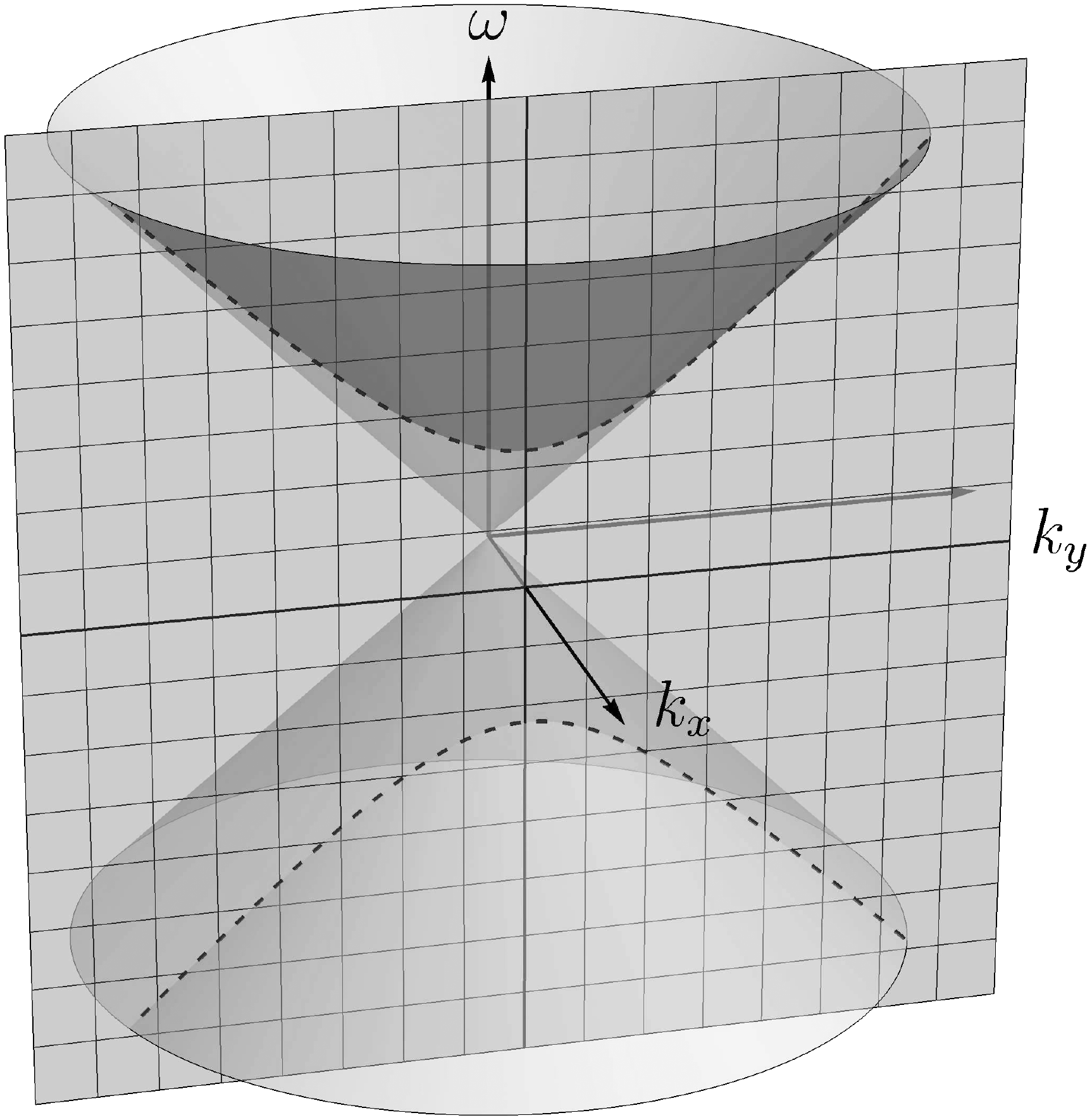}\label{disp-rel}}
\hspace{5mm}
\subfigure[Representation of the transmission of a relativistic massless particle through a square potential barrier step of height $V$. For an oblique collision the effective dispersion relation corresponds to a hyperbola. The dashed regions are the forbidden ones and separate the positive kinetic energy states (CB particles) from the negative ones (VB particles). For CB particles the momentum (long vector) points in the same direction as that of the velocity (short vector); for VB particles the momentum and the velocity point in opposite directions. For a frontal collision the hyperbola would degenerate in cone and there would be no forbidden regions. See~\cite{Calogeracos:1999yp,Allain:2011wk}]{
\includegraphics[scale=.99]{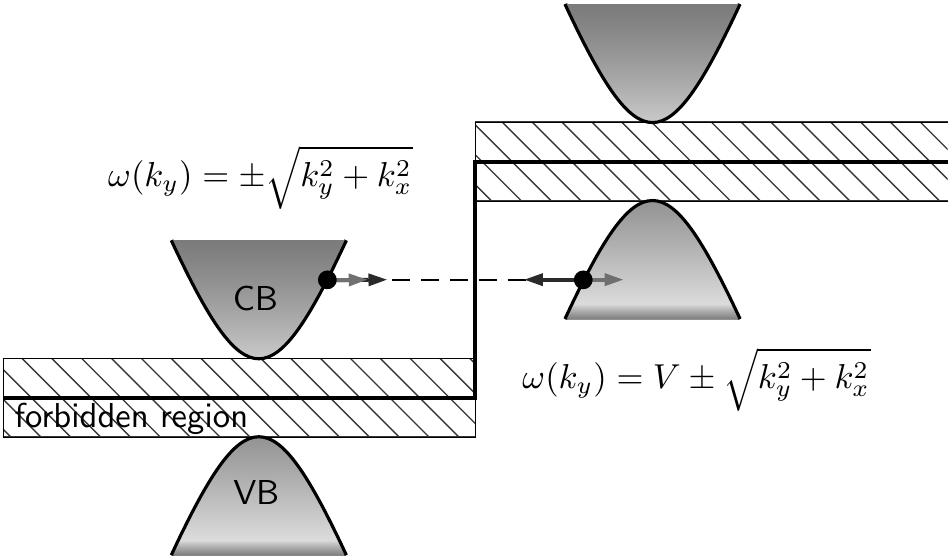}\label{CB-VB}}
\caption{Massles particle kinematics.} \label{disp-step}
\end{figure}
%%%%%%%%%%%%%%%%%%%%%%%%%%%%%%%%%%%%%%%%%%%%%%%
with $k_y=\sqrt{(\om-V)^2-k_x^2}$, 
if $(\om-V)^2>k_x^2$ (see Fig. \ref{disp-rel}). If $(\om-V)^2<k_x^2$,
the waves are real exponentials, hence do not propagate:
\eq
f^V_{\pm}(y) = e^{\pm \kappa y}, \quad
g^V_{\pm}(y) = -i\dfrac{k_x \pm \kappa}{\om-V} e^{\pm\kappa y},
\eqn{right-free-exp-waves}
with $\kappa =\sqrt{k_x^2-(\om-V)^2}$.

In the propagating case \equ{right-free-waves} the (unnormalized) 
probability density and 
fluxes \equ{density-and-current} are explicitly given by
\eq
\rho_\pm=2,\quad J_\pm^x=\dfrac{2k_x}{\om-V},\quad J_\pm^y
=\pm \dfrac{2k_y}{\om-V},
\eqn{fluxes-free}
which leads to the mean velocities (see \equ{local-velocity})
\eq
v_\pm^x=\dfrac{k_x}{\om-V},\quad v_\pm^y=\pm \dfrac{k_y}{\om-V},\quad \mbox{with}\ \
(v_\pm^x)^2+(v_\pm^y)^2=1.
\eqn{mean-velocity}
 Note that, thinking in the context of condensed matter, 
for $\om<V$, \ie for a negative kinetic energy, the propagation 
may be considered as 
of a particle in the (non-full) valence band (VB), whereas 
it is of a particle 
of positive kinetic energy in the (non-empty) conduction 
band (CB) if  $\om>V$.
In the former case, the direction of 
the flux, equal to that of the mean velocity, is opposed to that of the
phase velocity, the latter being proportional to $\mathbf{k}$.
This means that a right (left) mode as defined above corresponds 
in fact to a left (right) moving particle. 

Similiar considerations can be
made in the case of the propagation of holes. 

%%%%%%%%%%%%%%%%%%%%%%%%%%%%%%%%%%%%%%%%%%%%%%%%%%%%%%%%%
\subsubsection{Scattering boundary conditions}\label{scatt-conditions}

The free particle solutions \equ{right-free-waves} and \equ{right-free-exp-waves}, 
with $k_y$ and $\kappa$ substituted by
\eq 
k_{y\rm L,R} =\sqrt{(\om-V_{\rm L,R})^2-k_x^2},\quad
\kappa_{\rm L,R} =\sqrt{k_x^2-(\om-V_{\rm L,R})^2},
\eqn{q-kappa}
will be used in the following for 
the prescription of the asymptotic behaviour of the solutions of 
the Dirac equations \equ{component-Dirac-eq} in the cases of potentials 
obeying the condition \equ{compact-V}.
Interested in scattering states, we choose
boundary conditions such that we have a pure right moving particle 
state in the  
right asymptotic region $y\ge\yR$. In the case of a  CB state,
these conditions  will be taken as
\eq 
f({\bar y}) =  f^{V_{\rm R}}_{+}({\bar y}),\quad 
g({\bar y}) =  g^{V_{\rm R}}_{+}({\bar y}),
\eqn{bound-cond}
for the wave functions $f$, $g$ solutions of the interacting Dirac equation,
where ${\bar y}\ge\yR$ is some normalization point chosen in the right 
asymptotic region, $f_{+}^{V_{\rm R}}$ and $g_{+}^{V_{\rm R}}$ are the asymptotic wave functions 
defined in \equ{right-free-waves}, with $k_y=k_{y{\rm R}}$.  
For a  VB state, 
one has to substitute the index $+$ by the index $-$. 
These conditions  correspond to a process 
consisting of an incoming particle
coming from the left region $y\le\yL$: hence, in the right region $y\ge\yR$, 
one admits only the solution with flux pointing to the right -- hence a positive 
wave vector component $k_y$ for an outgoing CB particle and a negative one for 
a VB particle. 
In the left region both directions (incoming and reflecting) are allowed.
Again we have to distinguish the motions of a CB or of a VB particle. The proper distinction between CB and VB particles is crucial for the correct solution of the Klein 
phenomenon~\cite{stander:2009, Dragoman_2008, Katsnelson_2006, Calogeracos:1999yp, Das_2008, Katsnelson:2012, Thaller:1992}, 
 as mentioned in the Introduction.

The above holds for $k_{y{\rm R}}$
real. If, on the other hand, $k_{y{\rm R}}$ is imaginary, one has no propagating state in the right region. Therefore the boundary condition must select 
from \equ{right-free-exp-waves} the exponentially decreasing solution
to be put in the boundary condition:
\eq 
f({\bar y}) =  f^{V_{\rm R}}_{-}({\bar y}),\quad g({\bar y}) =  g^{V_{\rm R}}_{-}({\bar y}),
\eqn{bound-cond-exp-waves}
with $\kappa=\kappa_{\rm R}$ (see \equ{q-kappa}).

The conditions \equ{bound-cond} or  \equ{bound-cond-exp-waves} 
define uniquely the 
solution of the   Dirac equation with interaction.

%%%%%%%%%%%%%%%%%%%%%%%%%%%%%%%%%%%%%%%%%%%%%%%%%%%%%%%%%
\subsubsection{Reflection and transmission coefficients}

The reflection and transmission probabilities $\RR$ and $\TT$ are given,
in the present case of an $x$-independent potential, by the 
expressions
\eq 
\RR=\dfrac{|J^y_{{\rm L}-}|}{|J^y_{{\rm L}+}|},\quad 
\TT=\dfrac{|J^y_{{\rm R}+}|}{|J^y_{{\rm L}+}|},
\eqn{R&T}
where $J^y_{{\rm L}+}$, $J^y_{{\rm L}-}$ and $J^y_{{\rm R}+}$ 
are the $y$-components of
the incoming, reflecting and outgoing probability fluxes, 
respectively, in the asymptotic region. Note that 
$J^y_{{\rm R}-}=0$ in our scattering setting. Remember 
that, if $\om-\VL<0$, respectively $\om-\VR<0$, 
we have a VB particle propagating 
and the flux is in the direction opposed to that of the wave 
vector, in the left, respectively right, region.
Due to the $t$ and $x$ independence of the potential, the continuity equation
for the density and flux   reads $dJ_y(y)/dy=0$, hence 
$J^y_{{\rm L}+}$ + $J^y_{{\rm L}-}$
= $J^y_{{\rm R}+}$, which ensures the probability conservation 
$\RR+\TT=1$. 
 Note that in the gap $\VR-|k_x|<\om<\VR+|k_x|$
there is full opacity: $R=1$, $T=0$.

The fluxes are calculated according to \equ{density-and-current} 
in terms of the 
incoming, reflecting and outgoing wave functions obtained from the Fourier
coefficients of the wave function 
component\footnote{We could as well choose the other 
component, $g(y)$, without changing the result.}  $f(y)$ in the 
corresponding asymptotic regions (where the waves are free ones, 
the potential being constant):
 \eq\ba{l}
a_{{\rm L}\pm}=
\dfrac{k_{y{\rm L}}}{2\pi}
\dint_{\!\!\!\!\yL-2\pi/k_{y{\rm L}}}^{\yL} dy\, 
e^{\mp ik_{y{\rm L}}y}f(y),\quad%\\[6mm]
a_{{\rm R}\pm}=
\dfrac{k_{y{\rm R}}}{2\pi}\dint^{\yR+2\pi/k_{y{\rm R}}}_{\!\!\!\!\yR} dy\, e^{\mp ik_{y{\rm R}}y}f(y).
\ea\eqn{Fourier-coef} 
Let us calculate, as an example, the flux  $J^y_{{\rm L}+}$
for the incoming mode (in the region
 $y\le\yL$), in the case of a CB particle, \ie with $\om>\VL$). 
 The relevant spinor components are given by \equ{right-free-waves}, with ``$\pm$'' 
 substituted by ``$+$'' and multiplied by the Fourier coefficient $a_{{\rm L}+}$. This flux then is given by the last of Eqs. \equ{fluxes-free}, but multiplied by $|a_+|^2$. Doing the same for the other fluxes, we obtain the result (for incoming and outgoing  
 \textbf{CB particle} states 
\eq
J^y_{{\rm L}\pm}=\pm |a_{{\rm L}\pm}|^2 \dfrac{2k_{y{\rm L}}}{\om-\VL},\quad
J^y_{{\rm R}\pm}=\pm |a_{{\rm R}\pm}|^2 \dfrac{2k_{y{\rm R}}}{\om-\VR},
\eqn{fluxes-particle}
observing that $a_{{\rm R}-} = J^y_{{\rm R}-}=0$ due to the scattering
boundary conditions\equ{bound-cond}. For incoming or/and outgoing 
\textbf{VB particle} states one has to substitute $a_{{\rm L,R}\pm}$ 
by
 $a_{{\rm L,R}\mp}$ in the right-hand side of the first or/and 
second of Eqs. \equ{fluxes-particle}.
This result allows then to compute the reflection and transmission 
coefficients \equ{R&T}.

Observe that these calculations concern scattering states characterized with
both $k_{y{\rm L}}$ and $k_{y{\rm R}}$ (see \equ{q-kappa}) being real numbers. 
We discard the case 
of $k_{y{\rm L}}$ being imaginary, since this would mean the absence of 
an incoming mode. On the other hand, if  $k_{y{\rm R}}$ turns out to be 
imaginary, the boundary condition \equ{bound-cond}  selects the  solution 
which is exponentially decreasing in the region $y\ge\yR$. In this case, 
obviously, $\TT=0$ and $\RR=1$.

Bound states characterized by $k_{y{\rm L}}$ and $k_{y{\rm R}}$ both being imaginary
-- such  the bound states of a particle inside a potential well -- 
will not be   considered  in this paper.

%%%%%%%%%%%%%%%%%%%%%%%%%%%%%%%%%%%%%%%%%%%%%%%%%%%%%%%%%
\subsubsection{Comparison with the classical motion}\label{class-quantum-comp}
The relevant quantities which can thus be calculated are, beyond 
the wave functions, the fluxes \equ{density-and-current}, 
the reflection and transmission coefficients $\RR$ and $\TT$, 
and the mean velocities \equ{local-velocity} or, after integration, the
mean trajectories.
As a check of the ``correspondence principle'', we can compare the quantum
mean velocities and trajectories with the ones
obtained from the classical theory.

Classical velocities and trajectories are solutions of the equations of motion \equ{class-constr}  
and \equ{class-eq-E} of  Appendix \ref{app-electrostatic field} with the appropriate 
boundary conditions 
\eq 
x_{\rm cl}({\bar t}) = x_{\rm q}({\bar t}) = {\bar x},\quad
y_{\rm cl}({\bar t}) = y_{\rm q}({\bar t}) = {\bar y},\quad
\dot{x}_{\rm cl}({\bar t})=\bar{\dot{x}} = v^x_{\rm q}({\bar t}),
\eqn{traj-bound-cond}
taken at some time ${\bar t}$. $({\bar x},\,{\bar y})$ is some suitable
normalization  point, with ${\bar y}\le\yL$ if a reflection mode is considered, or ${\bar y}\ge\yR$ in the case of a transmission mode.
$\bar{\dot{x}} = v^x_{\rm q}({\bar t})$ is the $x$ component of the
mean quantum velocity at time $\bar t$.
%, ${\bar y}$ being the normalization value of $y$ used in the 
%quantum boundary conditions \equ{bound-cond}. 
The indices ``${{\rm cl}}$''
and ``${{\rm q}}$'' refer to classical quantities and quantum mean values, respectively.

%%%%%%%%%%%%%%%%%%%%%%%%%%%%%%%%%%%%%%%%%%%%%%%%%%%%%%%

\subsection{Some examples}

\begin{figure}[htb]
\centering
\includegraphics[scale=0.56]{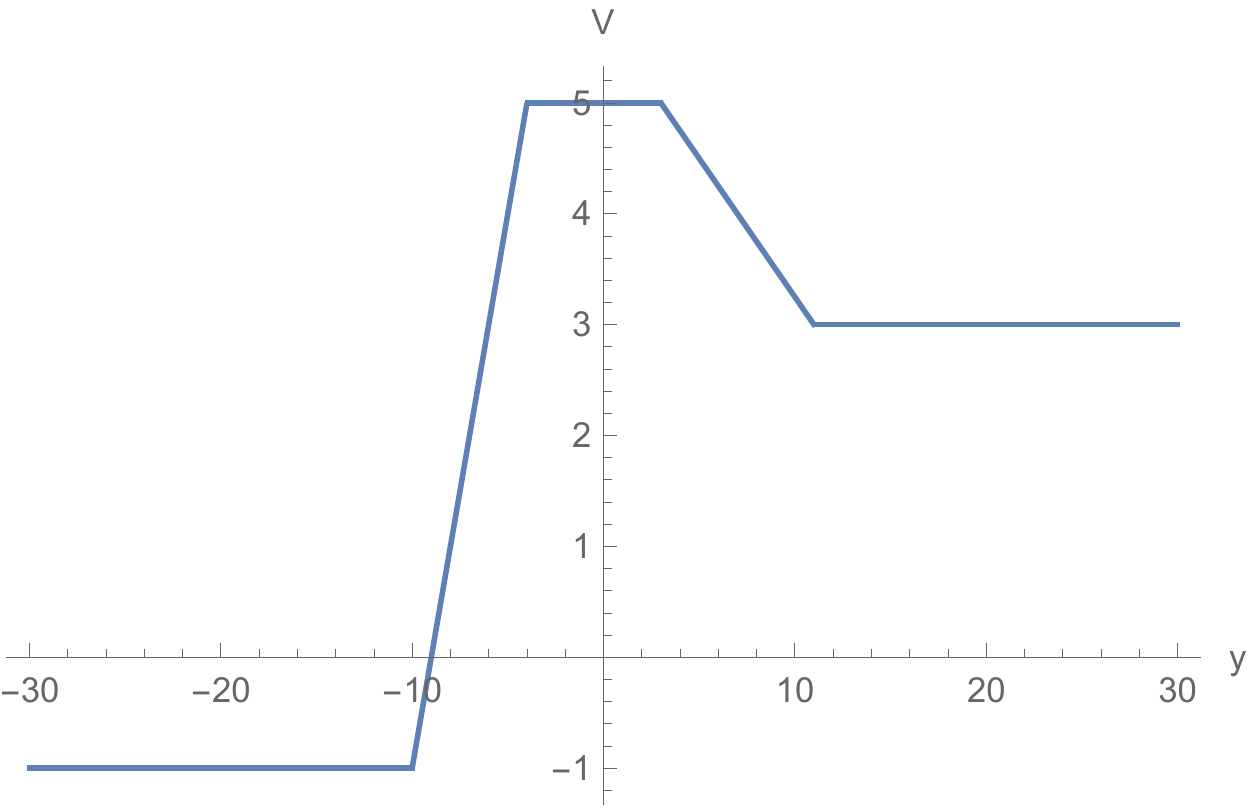}
\hspace{5mm}
\includegraphics[scale=0.56]{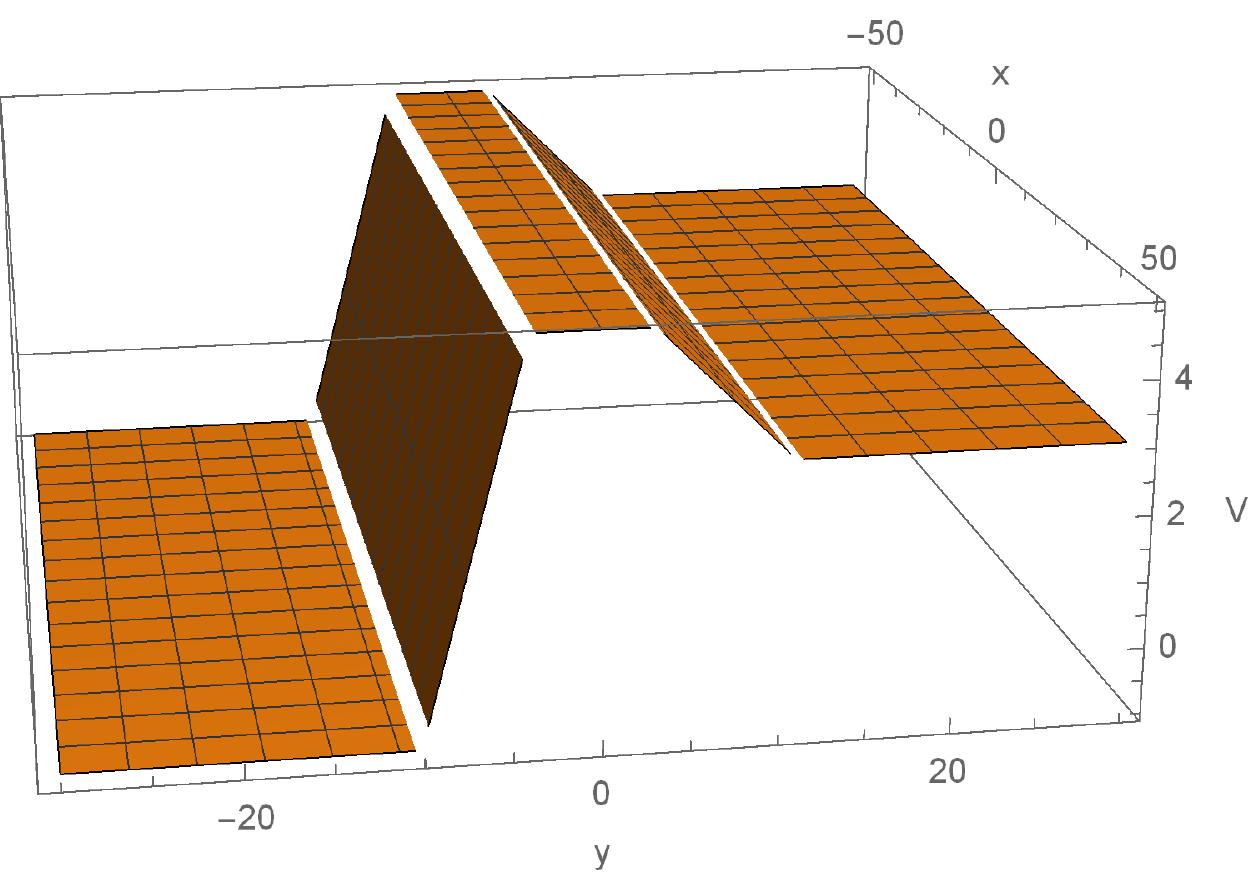}
\caption{\small 
Generic potential with parameters
$\{\yL,\,\yL',\,\yR',\,\yR\}$
= $\{-10,\,-4,\,3,\,11\}$ and $\{\VL,\,V_0,\,\VR\}$
= $\{-1,\,5,\,3\}$ in arbitrary units.
A 3D picture is shown, too.}
\label{potgeneric}
\end{figure}

Results  for some particular potentials are  presented in this Section. These potentials are  
of the square type or, more generally, 
piece-wise continuous functions $V(y)$ of the form (See Fig. \ref{potgeneric})
\eq
V(y) = \left\lbrace\ba{ll} 
\VL,&\quad y\le y_{\rm L},\es
\dfrac{V_0-\VL}{\yL'-\yL}(y-\yL)+\VL,
   &\quad \yL\le y\le \yL'\\[5mm]
V_0,&\quad \yL'\le y\le\yR'\es
\dfrac{\VR-V_0}{\yR-\yR'}(y-\yR)+\VR,
   &\quad \yR' \le y \le \yR\\[5mm]
V_{\rm R},&\quad \yR \le y
\ea\right.
\eqn{generic-pot}
%%%%%%%%%%%%%%%%%%%%%%%%%%%%%%%%%%%%%%%%%%%%%%%
%%%%%%%%%%%%%%%%%%%%%%%%%%%%%%%%%%%%%%%%%%%%%%%
The electric field is oriented  in the $y$-direction, with its value
given by
$E=- \dfrac{V_0-\VL}{\yL'-\yL}$ in the interval 
$(\yL, \yL')$, by $E=- \dfrac{V_0-\VR}{\yR'-\yR}$ in the interval
$(\yR',\yR)$ and by $E=0$ outside.

The potential \equ{generic-pot} will be substituted by a 
smoothed one in the numerical applications,
in order to avoid problems caused by the singularities at 
$\yL,\,\yL',\,\yR'$ and $\yR$.

In each of the examples shown below, the Dirac equation is solved using the scattering
 boundary conditions \equ{bound-cond} or \equ{bound-cond-exp-waves}
  explained in Subsection \ref{scatt-conditions}: the incoming wave describes a 
 particle emitted from the left half plane $y\le\yL$  (the left region with flat potential), 
  producing a reflected wave
to the left  and a transmitted wave to the right
describing the transmitted particle -- or, depending 
on the energy and momentum parameters, an exponentially 
decreasing wave corresponding 
to full opacity of the potential step or barrier.

%%%%%%%%%%%%%%%%%%%%%%%%%%%%%%%%%%

%%%%%%%%%%%%%%%%%%%%%%%%%%%%%%%%%%
\subsubsection{Square step potential}\label{Sq_step}

This is a slight generalization to 2 dimensions of the 
one-dimensional potential step problem 
found in the standard 
literature~\cite{Calogeracos:1999yp, Katsnelson:2012, Das_2008} ,
with the $y$-dependent potential
\eq
V(y) = \left\lbrace\ba{ll} 
0,&\quad y< 0,\es
V>0,&\quad y> 0,
\ea\right.
\eqn{square-step}
The solution of the Dirac equation as an eigenvector of the energy 
with value $\om$, and of the 
$x$-component of the linear momentum 
with value $k_x$ (see \equ{Dirac spinor}),
and with the scattering boundary condition defined in
Subsection \ref{scatt-conditions},  is given, in the case of a
 \textbf{particle} in both sides, \ie with $\om>V$, by\footnote{ Recall that the suffixes $+$ and $-$ refer to the sign of the phase velocity as defined in 
 Eqs. \equ{right-free-waves}.}
\eq
 \Psi(t,x,y)=\lac \ba{ll}
 e^{-i\om t+i k_x x} \lp\ba{c} f^{\rm L}_+(y) + A f^{\rm L}_-(y)\\ 
                             g^{\rm L}_+(y) + A g^{\rm L}_-(y)\ea\rp
 &\quad (y<0),\\[5mm]
e^{-i\om t+i k_x x} \lp\ba{c}B f^{\rm R}_-(y)\\ 
                           B g^{\rm R}_-(y)\ea\rp
 &\quad (y>0),\ea\right.
\eqn{sol-square-step}
where $A$ and $B$ are coefficients fixed by the continuity condition
\[
\Psi(t,x,y)|_{y=-0} = \Psi(t,x,y)|_{y=+0},
\]
with the result\footnote{In this and the next subsection, 
we consider a massive particle for the sake of comparison 
with the literature~\cite{stander:2009, Dragoman_2008, Katsnelson_2006, Calogeracos:1999yp}.}
\eq
A= \lac\ba{ll} \dfrac{k_{y{\rm L}}(\om-V-m) -k_{y{\rm R}}  (\om-m) - ik_xV}
{k_{y{\rm L}}(\om-V-m) + k_{y{\rm R}}  (\om-m) + ik_xV},
&\quad \om<0,\\[5mm]
\dfrac{-k_{y{\rm L}}(\om-V-m) + k_{y{\rm R}}  (\om-m) - ik_xV}
{k_{y{\rm L}}(\om-V-m) + k_{y{\rm R}}  (\om-m) + ik_xV},
&\quad 0<\om < V,\\[5mm]
\dfrac{k_{y{\rm L}}(\om-V-m) - k_{y{\rm R}}  (\om-m) + ik_xV}
{k_{y{\rm L}}(\om-V-m) + k_{y{\rm R}}  (\om-m) - ik_xV},
&\quad \om>V,
\ea\right.
\eqn{coeffA-sq-step}
\eq
B= \lac\ba{ll} \dfrac{2k_{y{\rm L}}(\omega-m)}
{k_{y{\rm L}}(\om-V-m) + k_{y{\rm R}}  (\om-m) + ik_xV},
&\quad \om<0,\\[5mm]
\dfrac{2k_{y{\rm L}}(\omega-m)}
{k_{y{\rm L}}(\om-V-m) - k_{y{\rm R}}  (\om-m) - ik_xV} ,
&\quad 0<\om < V,\\[4mm]
\dfrac{2k_{y{\rm L}}(\omega-m)}
{k_{y{\rm L}}(\om-V-m) + k_{y{\rm R}}  (\om-m) - ik_xV},&\quad \om>V,
\ea\right.\\[4mm]
\eqn{coeffB-sq-step}
where the $y$-components $k_{y{\rm L}}$ and $k_{y{\rm R}}$ of the 
wave vector  are given
by \equ{q-kappa} with $V_{\rm L}=0$ and $V_{\rm R}=V$.

The reflection and transmission coefficient \equ{R&T} take then the form
\eq\ba{l}
\RR=|A|^2=\es
 \lac\ba{ll}
 \hspace{-2mm}\dfrac{\lp -k_{y{\rm L}}(\om-V-m)  +  k_{y{\rm R}}(\om-m)\rp^2 +k_x^2V^2}
{\lp k_{y{\rm L}}(\om-V-m)  +  k_{y{\rm R}}(\om-m)\rp^2 +k_x^2V^2},
&\quad \om<-\sqrt{k_x^2+m^2},\\[5mm]
 \hspace{-2mm}\dfrac{\lp k_{y{\rm L}}(\om-V-m)   + k_{y{\rm R}}(\om-m)\rp^2 +k_x^2V^2}
{\lp -k_{y{\rm L}}(\om-V-m)   + k_{y{\rm R}}(\om-m)\rp^2 +k_x^2V^2},
&\null\hspace{-1mm} \sqrt{k_x^2+m^2}<\om < V-\sqrt{k_x^2+m^2},\\[5mm]
 \hspace{-2mm}1,
&\hspace{-14mm}
\quad V-\sqrt{k_x^2+m^2}<\om < V+\sqrt{k_x^2+m^2},\es
 \hspace{-2mm}\dfrac{\lp -k_{y{\rm L}}(\om-V-m)   + k_{y{\rm R}}(\om-m)\rp^2 +k_x^2V^2}
{\lp k_{y{\rm L}}(\om-V-m)  +  k_{y{\rm R}}(\om-m)\rp^2 +k_x^2V^2},
&\quad  \om>V+\sqrt{k_x^2+m^2},
\ea\right.
\ea\eqn{R-sq-step}
and
\eq
\TT=|B|^2\,\dfrac{k_{y{\rm R}}(\om-V-m)}{k_{y{\rm L}}(\om-m)} =1-\RR \hspace{75mm} 
\eqn{T-sq-step}
We note that in the cases where $-\sqrt{k_x^2+m^2}<\om<\sqrt{k_x^2+m^2}$ 
or $V-\sqrt{k_x^2+m^2}<\om<V+\sqrt{k_x^2+m^2}$, 
the wave vector components $k_{y{\rm L}}$ or $k_{y{\rm R}}$, respectively,
 are imaginary, which corresponds to real exponential waves. 
 The first case is discarded since there is then no propagating
 incident particle. In the second case, there is no transmitted 
 propagating particle, hence the reflection probability $\RR$ is equal to 1.

One checks that $\RR+\TT=1$ , as it should.

One recovers the standard literature result~\cite{Calogeracos:1999yp} 
for the 1-dimensional system
by taking $k_x=0$ in \equ{R-sq-step}, \equ{T-sq-step}, \ie 
 a vanishing $x$-component of the wave vector.

Let us note, at this point, that the result, taken at $k_x=0$, does not coincide with the expression produced in part of the 
literature~\cite{stander:2009, Dragoman_2008, Katsnelson_2006, Calogeracos:1999yp}. 
The latter gives values for $\RR$ and $\TT$ outside of the interval $(0,1)$,
a fact called the ``Klein paradox``. As it is explained 
in~\cite{Calogeracos:1999yp} 
 this apparent paradox appears if one 
forgets that  VB particle propagation occurs in the Dirac theory 
at values of the energy for which the non-relativistic quantum 
theory would yield an exponential damping. This is what happens, in the 
present example, for the incident, reflected or transmitted object if 
$\om<-\sqrt{k_x^2+m^2}$, and 
for the transmitted one if
$\sqrt{k_x^2+m^2}<\om < V-\sqrt{k_x^2+m^2}$. On the other hand, there 
is exponential damping if $V-\sqrt{k_x^2+m^2}<\om<V+\sqrt{k_x^2+m^2}$.

\begin{figure}[htb]
\centering
\includegraphics[scale=0.45]{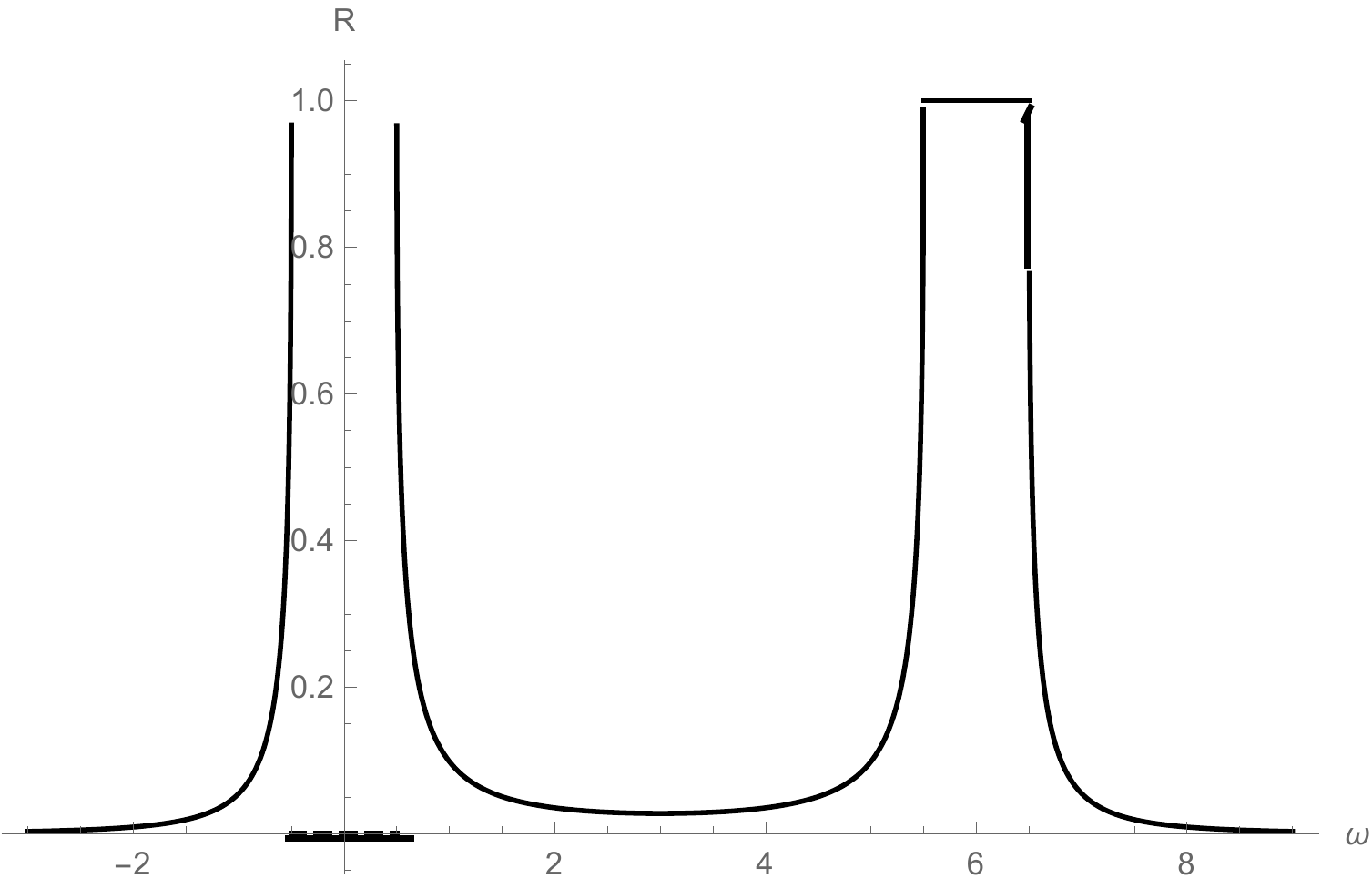}\hspace{4mm}
\includegraphics[scale=0.45]{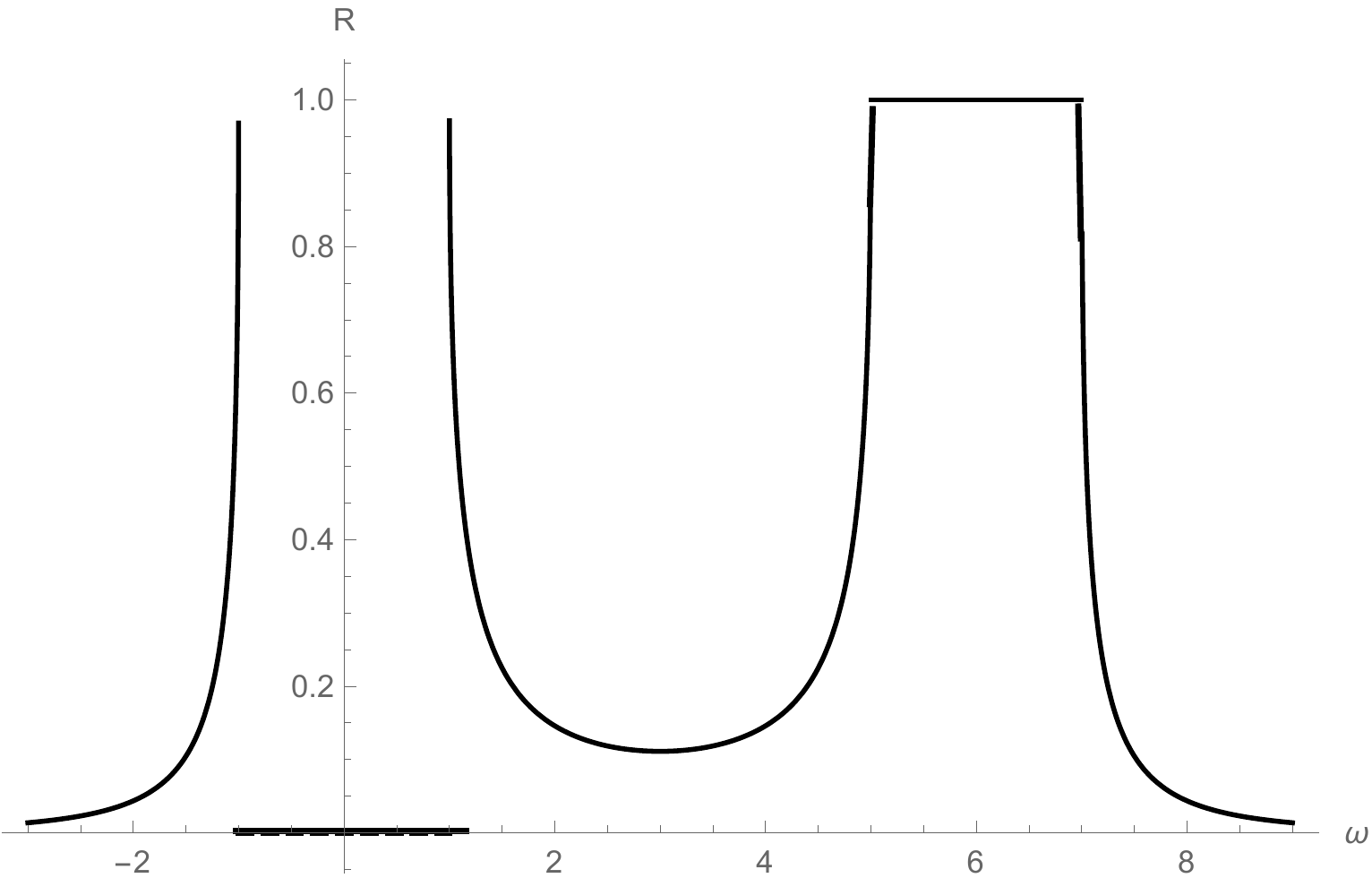}\hspace{4mm}
\includegraphics[scale=0.45]{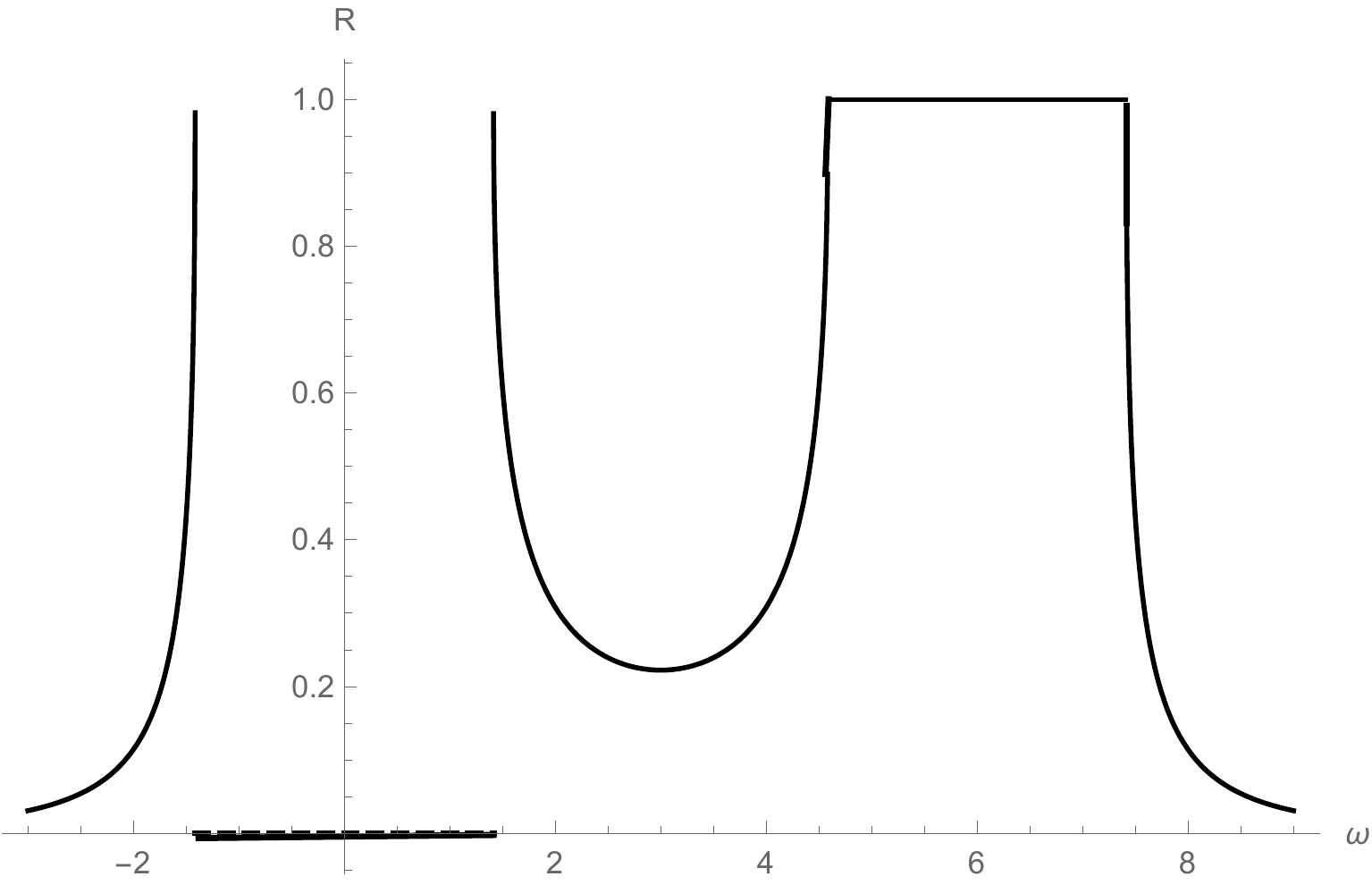}
\caption{\small Square step potential \equ{square-step} with $V=6$: 
reflection probability $\RR$
 as a function of the frequency $\om$ for ($m,k_x$)=
 ($0,0.5$), 
 $(0,1.0)$ and  ($1.0,1.0$), respectively. 
 The heavy horizontal segment shows the 
 energy gap interval..
}
\label{fig_sq-step}
\end{figure}

Fig. \ref{fig_sq-step} shows the behaviour of the reflection probability
as a function of the energy $\om$ for three sets of parameters' values. One observes an increase of the forbidden region and of the region of total reflection when either $|k_x|$ or $m$ increases.
%%%%%%%%%%%%%%%%%%%%%%%%%%%%%%%%%%%%%%%%%%%%%%%
%%%%%%%%%%%%%%%%%%%%%%%%%%%%%%%%%%

%%%%%%%%%%%%%%%%%%%%%%%%%%%%%%%%%%%%%%%%%%%%%%%%%%%%%%%
\subsubsection{Square barrier potential}\label{Sq barrier}

This is again a slight generalization to 2 dimensions of the 
 problem of the one-dimensional potential 
 barrier~\cite{Calogeracos:1999yp}, 
 with the 
$y$-dependent potential
\eq
V(y) = \left\lbrace\ba{ll} 
0,&\quad y< -a \mbox{ or }y>a\ (a>0),\es
V>0,&\quad -a<y<a,
\ea\right.
\eqn{square-barrier}
The solution of the (massive) Dirac equation 
as well as the calculation of the reflection and transmission 
probabilities $\RR$ and $\TT$ follow the same 
lines as for the potential step in the preceding subsection and will 
not be detailed here. The results for $\RR$ and $\TT$ happen to coincide with the solution found in~\cite{Calogeracos:1999yp} 
 for the 1-dimensional problem\footnote{This 
is not the case in the example of the step potential
examined in the preceding subsection, where $\RR$ and
$\TT$ depend on both independent variables $k_x$ and $m$.}, 
but with the 
mass parameter  $m$ substituted by $\sqrt{k_x^2+m^2}$:
\eq
\RR= \dfrac{(1-\la)^2\sin^2(2ak_y)}{4\la+(1-\la)^2\sin^2(2ak_y)}  ,
\quad\quad \TT = 1-\RR,
\eqn{RT-sq-barrier}
with $k_y=\sqrt{(\om-V)^2-k_x^2-m^2}$ and 
\[
\la =\dfrac{( V - \om+\sqrt{k_x^2 + m^2})(\om+\sqrt{k_x^2 + m^2})}
     {(V - \om-\sqrt{k_x^2 + m^2})
      (\om-\sqrt{k_x^2 + m^2}) },
\]
for $\om>\sqrt{k_x^2+m^2}$ (propagation of a CB particle) or  
$\om<-\sqrt{k_x^2+m^2}$ (propagation of an VB particle). 
These probabilities are not defined in the gap interval 
$-\sqrt{k_x^2+m^2}<\om<\sqrt{k_x^2+m^2}$, where there is no propagation at all.

\begin{figure}[htb]
\centering
\includegraphics[scale=0.35]{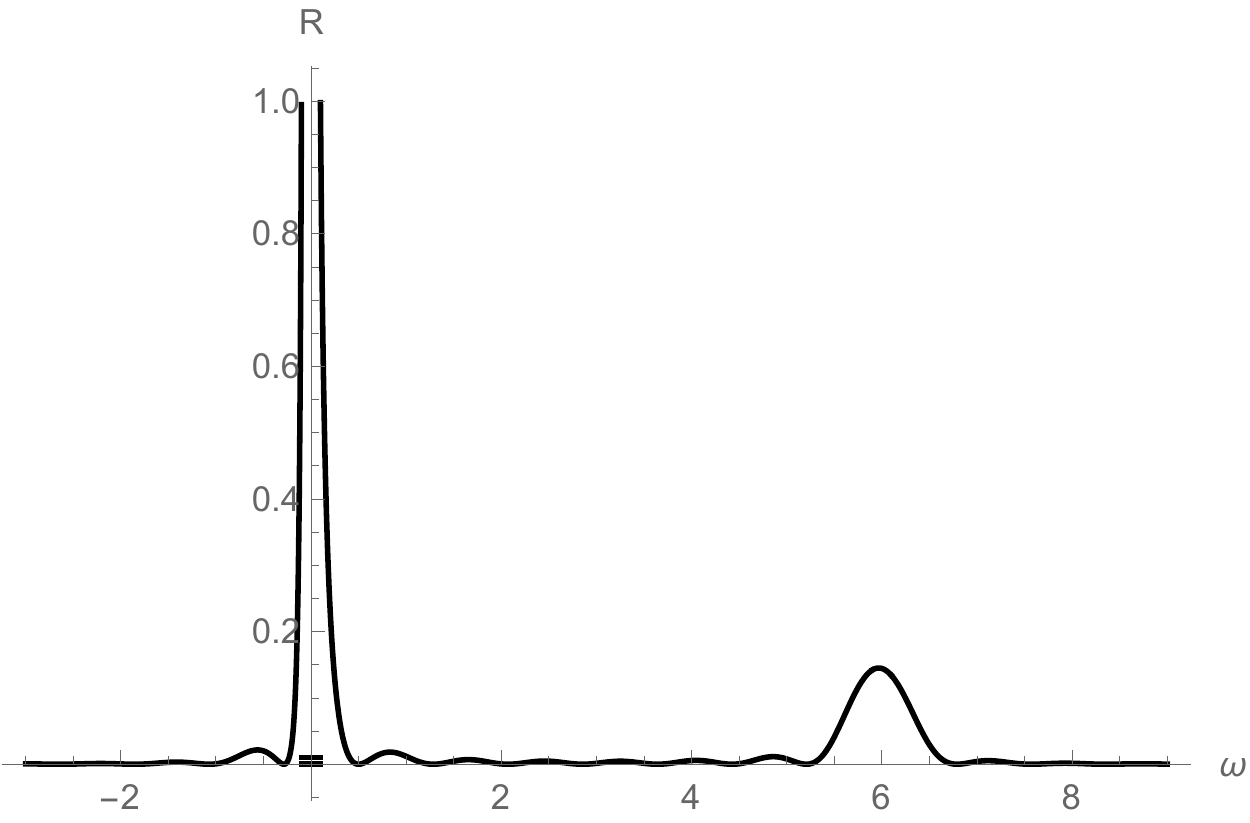}\hspace{4mm}
\includegraphics[scale=0.35]{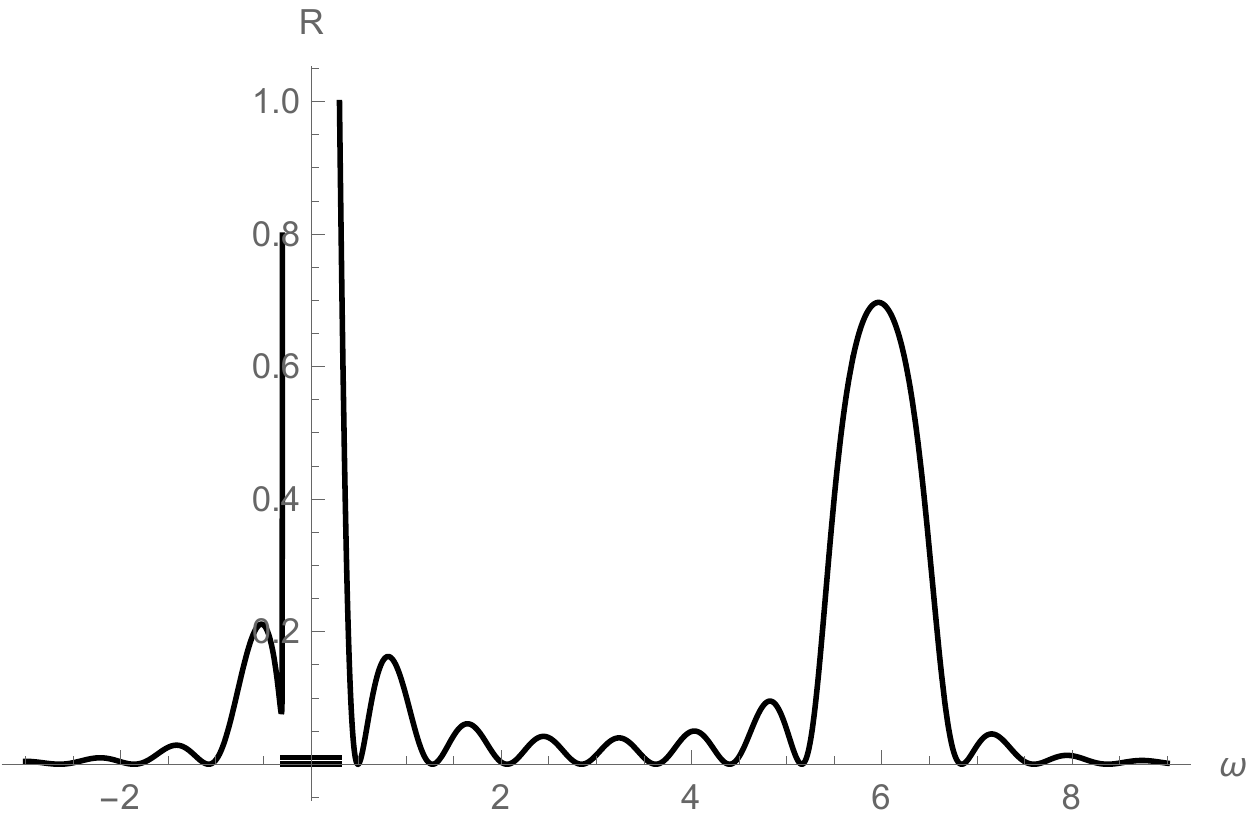}\hspace{4mm}
\includegraphics[scale=0.35]{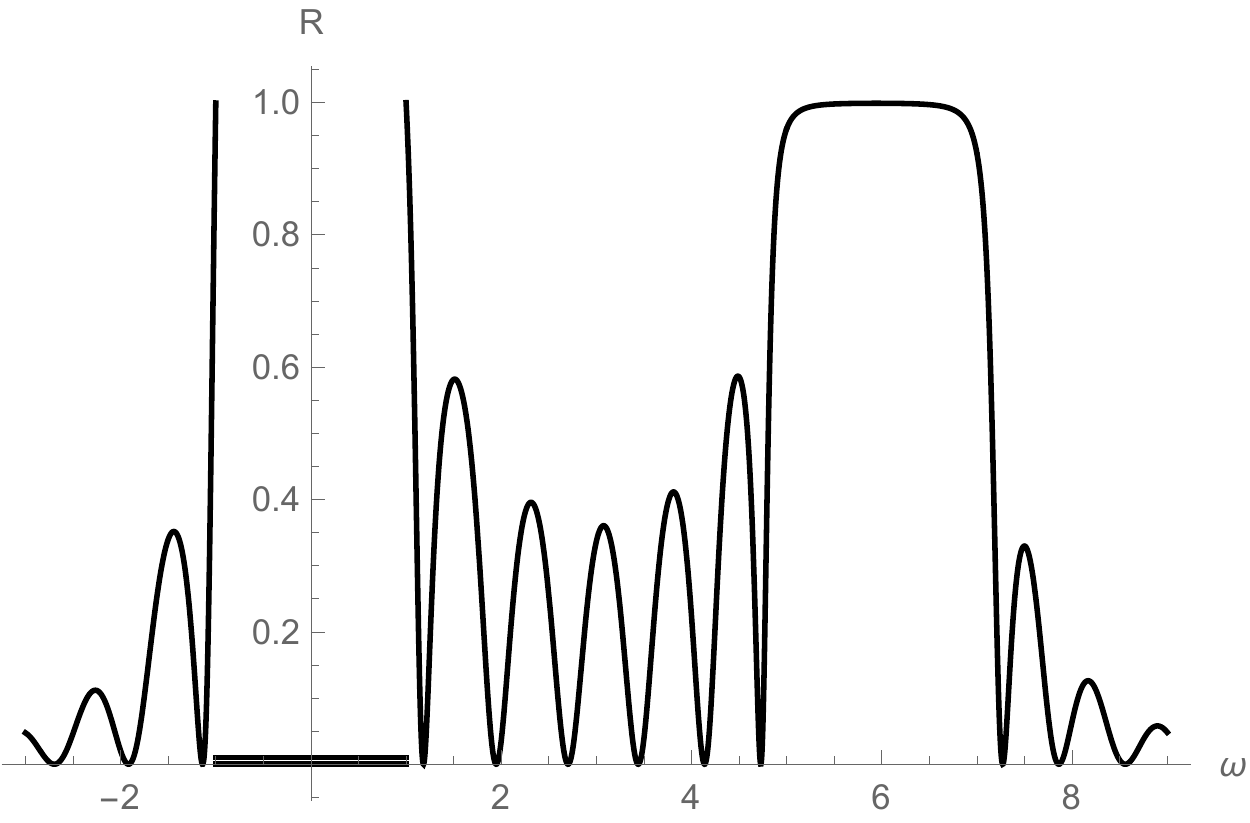}
%\hspace{5mm}
%\includegraphics[scale=0.45]{fig_sq-step_m=0_k_x=1.pdf}
\caption{\small Square barrier potential \equ{square-barrier}: Reflection probability $\RR$
 as a function of the frequency $\om$ in the massless case for $k_x=$
 $0.1$, 
 $0.3$ and  $1.0$, respectively. The heavy horizontal segment shows the  energy gap interval.  The potential parameters are $V=6$ and $a=5$.
}
\label{fig_sq-barrier}
\end{figure}

Fig. \ref{fig_sq-barrier} shows the behaviour of the reflection probability $\RR$
as a function of the energy $\om$ for three sets of parameters' values
in the massless case. One observes a decrease of the forbidden region 
when $|k_x|$ decreases. As can be seen from \equ{RT-sq-barrier}, for a vanishing 
momentum $x$-component, \ie for a frontal incidence, there is total transparency: $\RR=0$
for $k_x=0$.
 The oscillations  in the allowed region correspond to the so-called transmission 
 resonance phenomenon~\cite{Calogeracos:1999yp}. 
%%%%%%%%%%%%%%%%%%%%%%%%%%%%%%%%%%%%%%%%%%%%%%%
%%%%%%%%%%%%%%%%%%%%%%%%%%%%%%%%%%%%%%%%%%%%%%%

%%%%%%%%%%%%%%%%%%%%%%%%%%%%%%%%%
\subsubsection{Oblique step potential}\label{Obl step}

\begin{figure}[htb]
\centering
\includegraphics[scale=0.55]{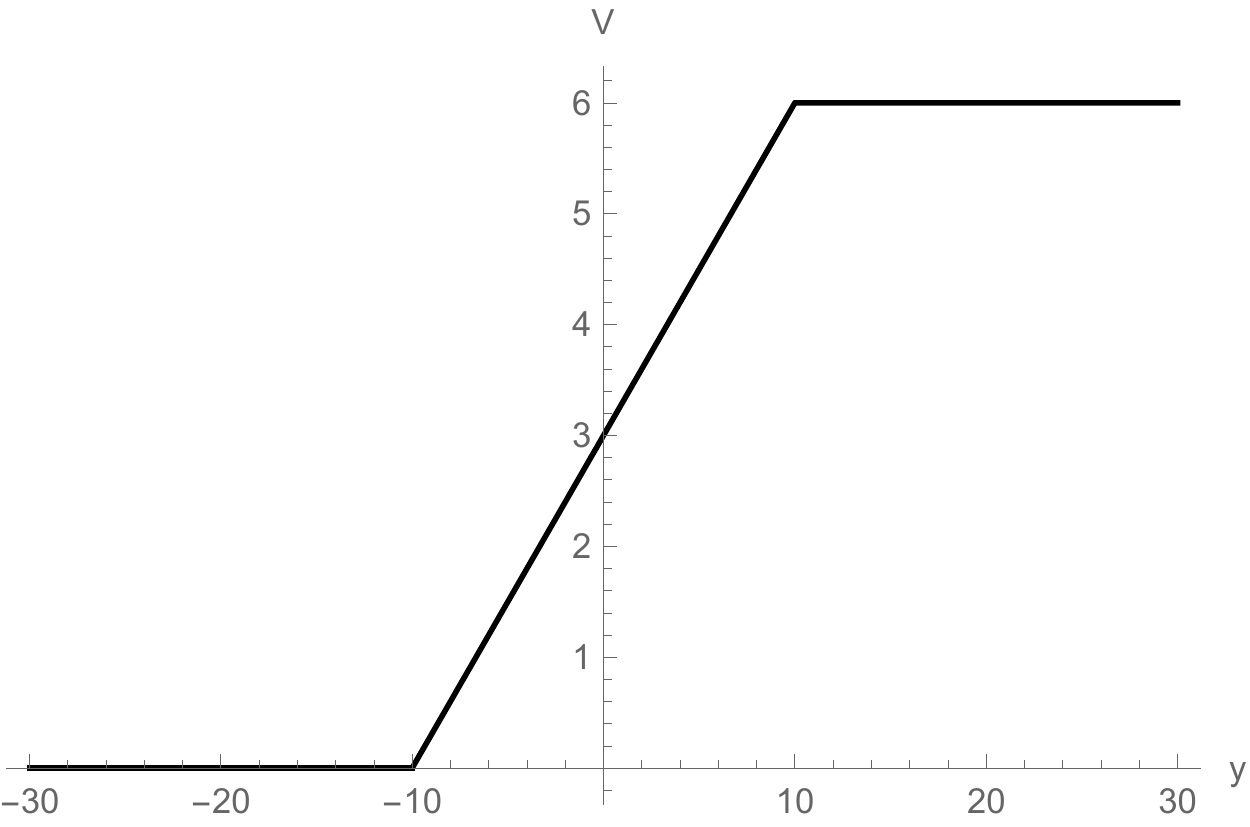}
\caption{\small 
 \hfill Oblique step potential. The values of the parameters of \equ{generic-pot} are taken as 
 $\yL=-10$, $\yL'=\yR'=\yR=10$, $\VL=0,\ V_0=\VR=6$.
}
\label{figstep}
\end{figure}

We consider here the stepwise potential $V(y)$,
a smoothed version of the one shown in Fig. \ref{figstep}. 
%%%%%%%%%%%%%%%%%%%%%%%%%%%%%%%%%%%%%%%%%%%%%%%
%%%%%%%%%%%%%%%%%%%%%%%%%%%%%%%%%%%%%%%%%%%%%%%
In this example and in the next ones, the massless 
Dirac equation \equ{Dirac-eq} as well as the dynamical quantities of interest are solved and calculated numerically using the software 
Mathematics~\cite{Wolfram}. 

\begin{figure}[htb]
\centering
\includegraphics[scale=0.41]{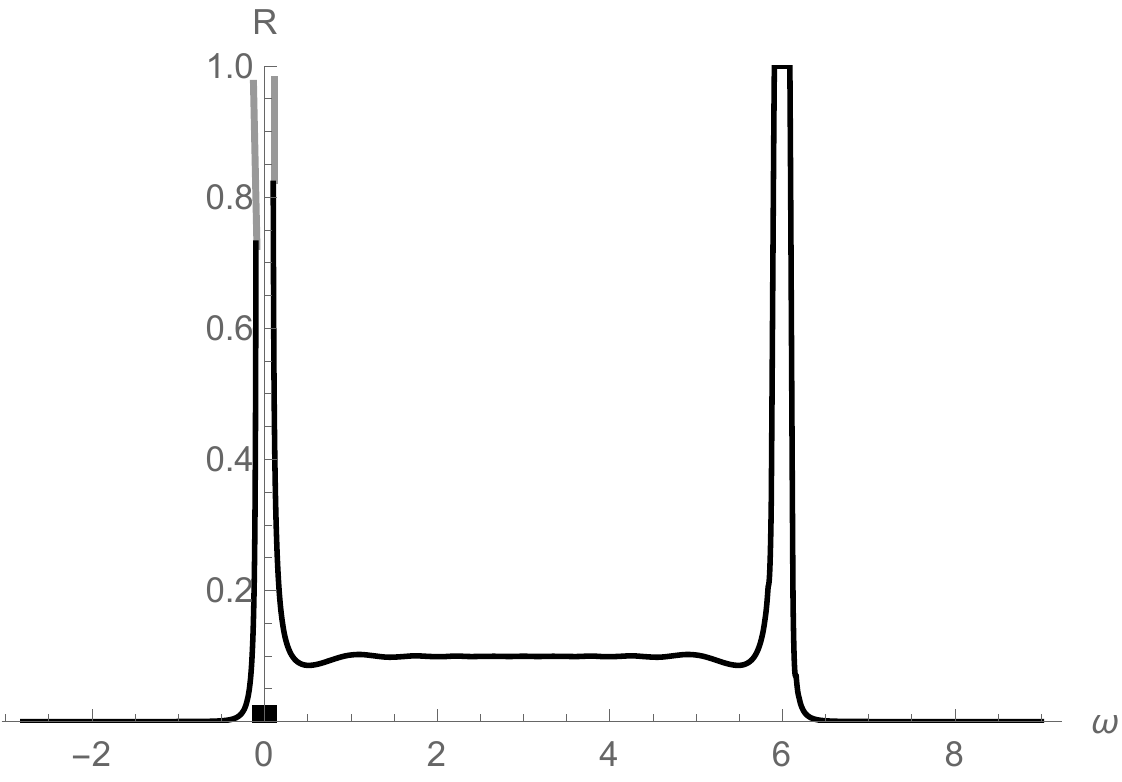}\hspace{4mm}
\includegraphics[scale=0.39]{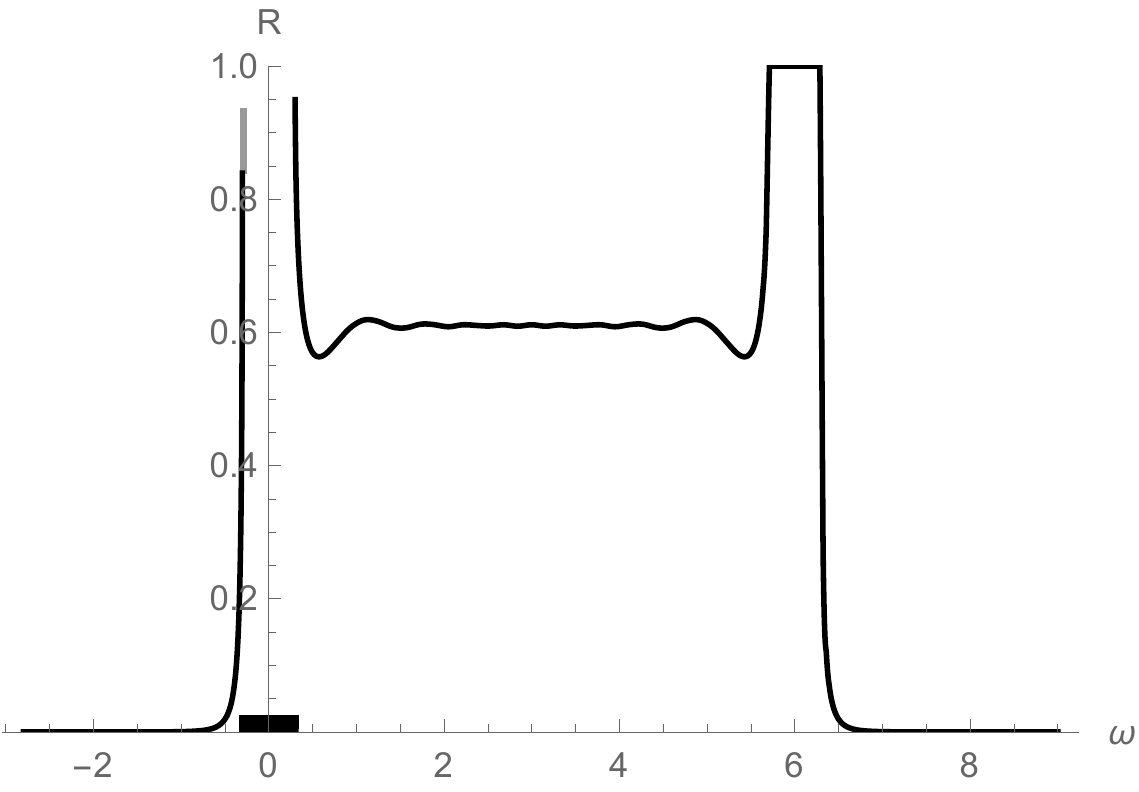}\hspace{4mm}
\includegraphics[scale=0.37]{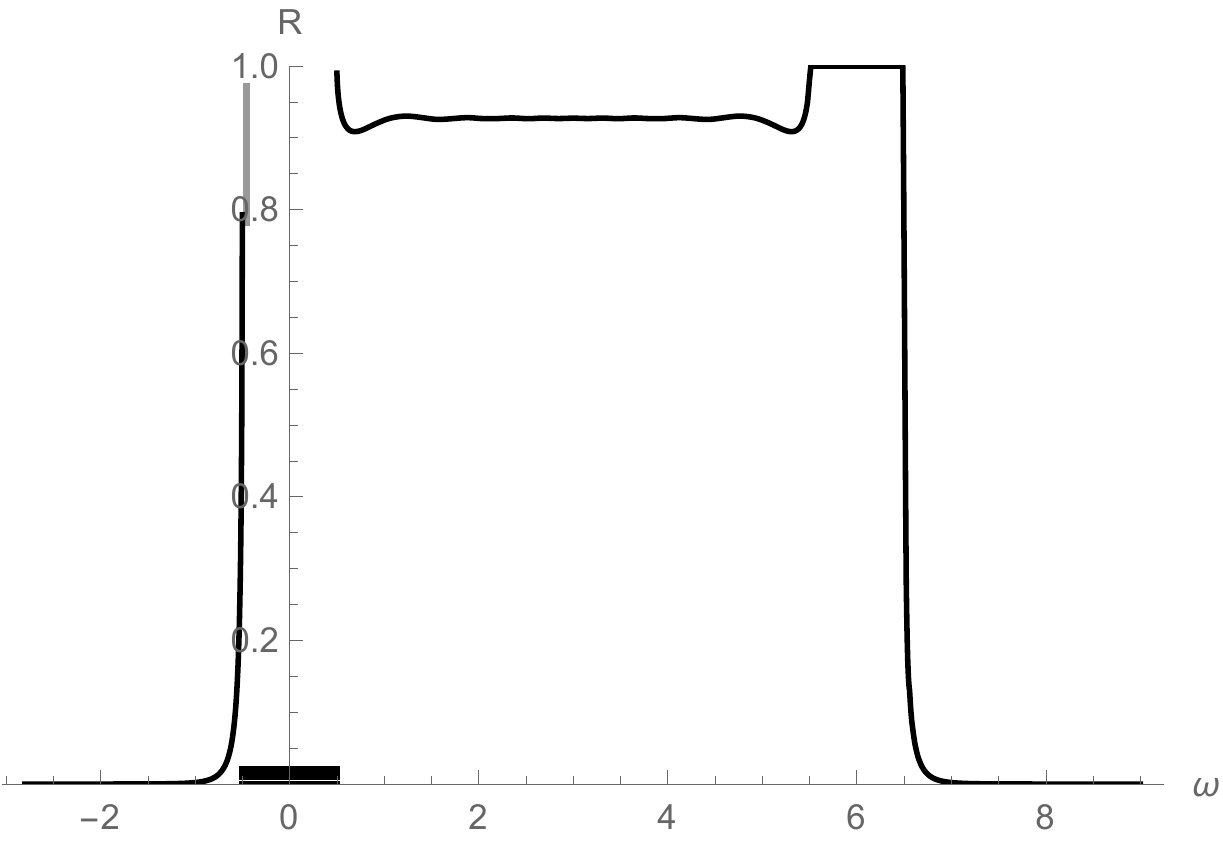}%\hspace{4mm}
\caption{\small 
 Potential of Fig. \ref{figstep}:  Reflection probability $\RR$
 as a function of the frequency $\om$ in the massless case for $k_x=0.1$, 
 $0.3$ and  $0.5$, respectively. The heavy horizontal segment shows the 
 energy gap interval.  
}
\label{fig_obl-step_R&T}
\end{figure}

Fig. \ref{fig_obl-step_R&T} shows the reflection probability $\RR$ as a
function of the energy $\om$ for various values of the $x$-component $k_x$
of the momentum.
%%%%%%%%%%%%%%%%%%%%%%%%%%%%%%%%%%%%%%%%%%%%%%%
%%%%%%%%%%%%%%%%%%%%%%%%%%%%%%%%%%%%%%%%%%%%%%%

One observes features very similar to those of the square step potential
seen in Subsect. \ref{Sq_step}.
Besides the expected energy gap, one recovers the 
 ``Klein phenomenon'':
 Complete opacity for energies in the region $\VR-|k_x|<\om<\VR+|k_x|$, and
appreciable transparency in the region 
$\VL+|k_x|<\om<\VR-|k_x|$ where opacity would be complete in the 
non-relativistic theory. Recall that
$\VL$ and $\VR$ are the values of the potential in the left and
right region, respectively. Also, as in the  square step case, 
the transparency tends to increase when the absolute 
value of $|k_x|$ decreases, being complete for $k_x=0$, \ie for an incident
wave vector orthogonal to the potential barrier.

\begin{figure}[htb]
\centering
\includegraphics[scale=0.36]{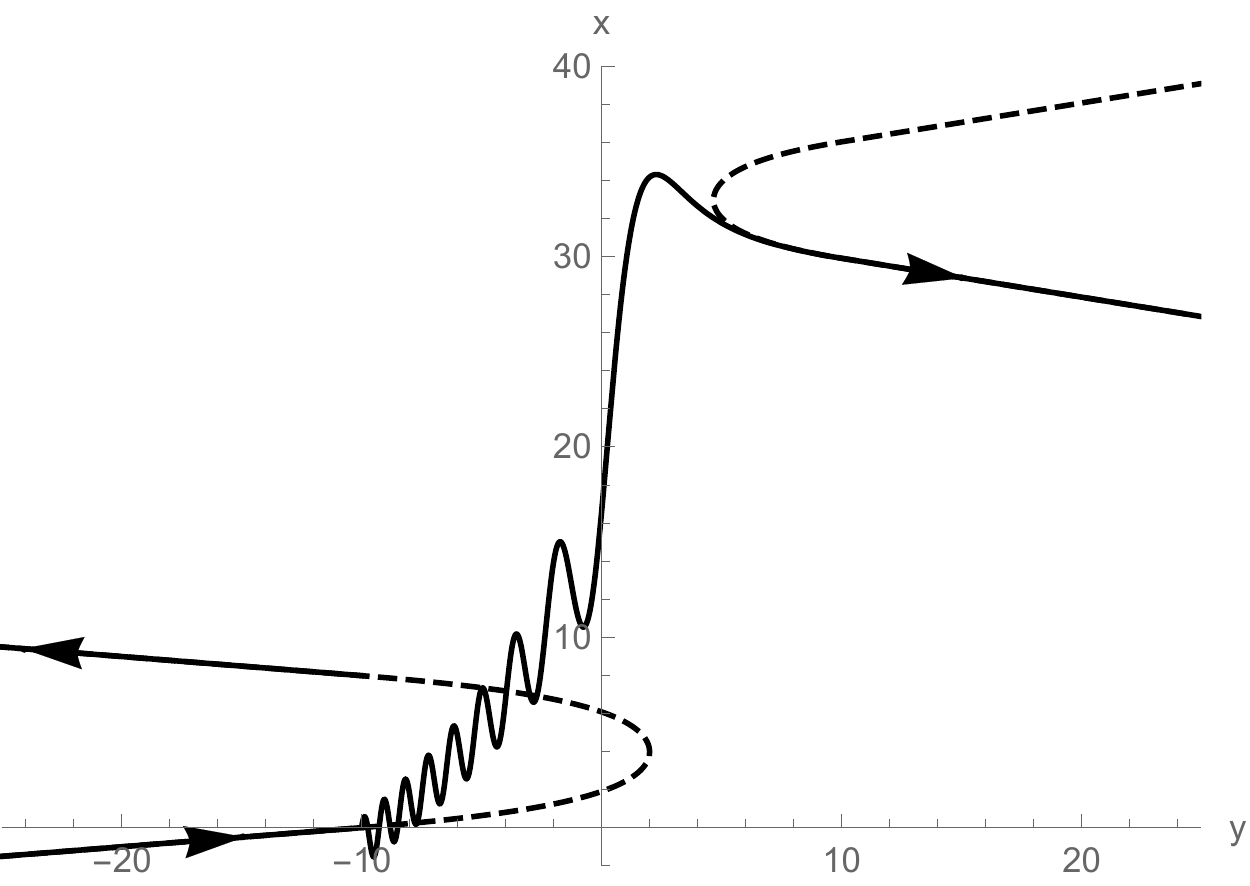}\hspace{4mm}
\includegraphics[scale=0.36]{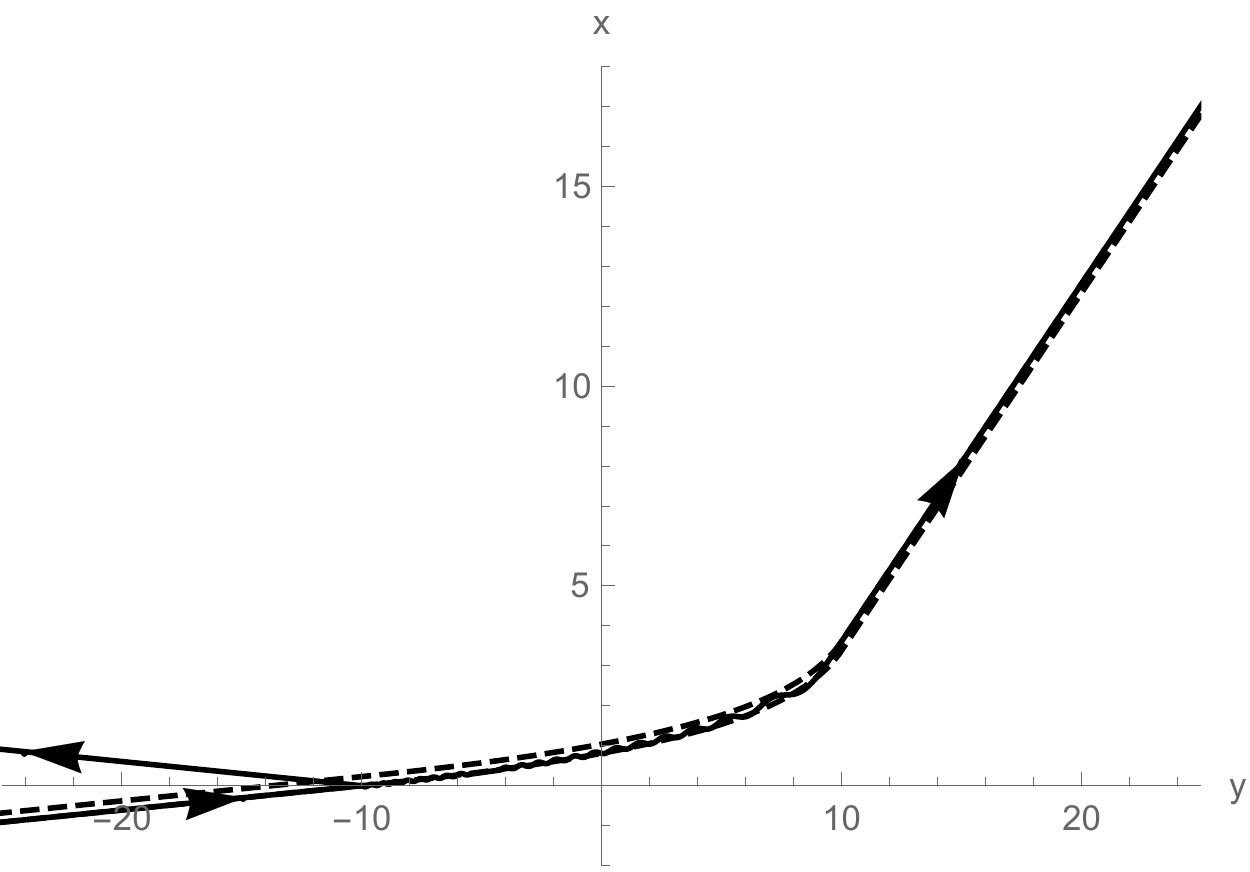}\hspace{4mm}
\includegraphics[scale=0.36]{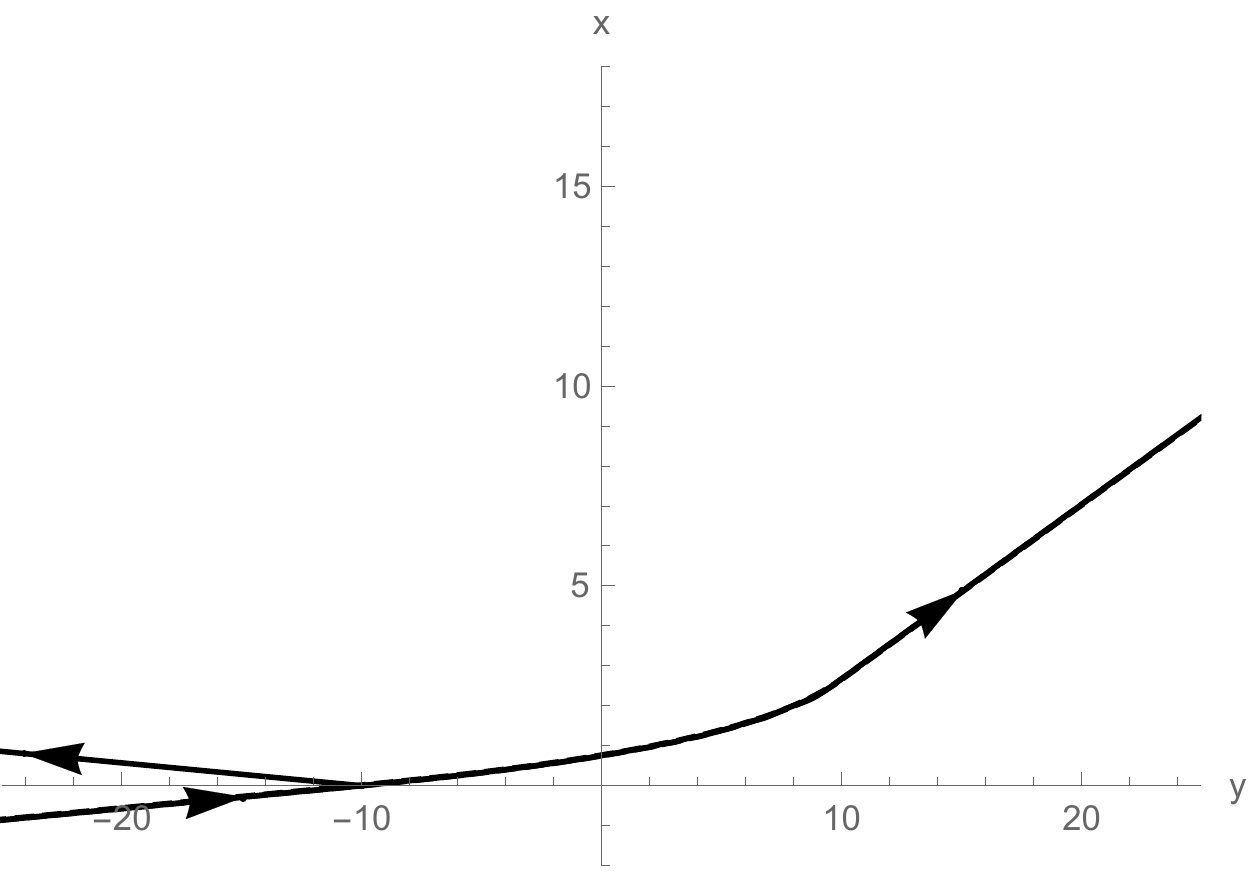}%\hspace{4mm}
\caption{\small 
 Potential of of Fig. \ref{figstep}: Quantum mean trajectories 
 (continuous lines) are
 shown together with the corresponding classical trajectories 
 (dashed lines), for $(\om,k_x)$ = $(4.0,0.4)$, $(6.6,0.4)$ and $(7.0,0.4)$.  
The respective values of the reflection probability $\RR$ are 0.812, 0.021 and  0.008.}
\label{fig_obl-step_traj}
\end{figure}

Fig. \ref{fig_obl-step_traj} shows the quantum mean trajectories compared with the corresponding classical ones for one value of $k_x$ and 
three values of the energy $\om$. 
We show in the left part of the graphics both the incident and reflected
particle quantum paths, with arrows indicating the direction of 
the mean velocity vector. In the right-hand part only the transmitted 
particle path appears, by construction, due to  
the boundary conditions corresponding
to an incident particle coming from the left.

%%%%%%%%%%%%%%%%%%%%%%%%%%%%%%%%%%%%%%%%%%%%%%%%%%%%%%%
%%%%%%%%%%%%%%%%%%%%%%%%%%%%%%%%%%%%%%%%%%%%%%%

The first case shown in Fig. \ref{fig_obl-step_traj} 
exemplifies the case of the energy 
lying between the bottom and
top values $(\VL,\VR)$ = $(0,6)$ of the potential, 
where the reflection is appreciable -- 
it would be total in the classical case. The incoming and reflection 
modes are those of a CB particle, whereas the transmitted one is that of
a VB particle.
On the classical level, there are 
corresponding trajectories both for the reflection 
of a CB particle coming from the left or for a VB particle 
coming from the right. Both are shown in the figure as dashed lines. 
We see that the quantum mean trajectories follow the classical paths
whenever there are given by pure left or right progressive waves, as
it is the case
outside of the interaction domain $(\yL,\yR)$ = $(-10,10)$. Inside
this domain, one sees a somewhat wild behaviour of the quantum trajectory 
-- a \textit{Zitterbewegung} effect due to the superposition of right moving and left moving waves. However, when $y$ approaches $\yR$ from below,
the trajectory becomes increasingly smooth and coincident 
with the classical one or, in other words, becomes a more and more pure
right moving mode.

The other two cases shown in Fig.  \ref{fig_obl-step_traj}, 
with small reflection 
probabilities, are typical scattering states, the energy being
above the top value of 
the potential, $\om>\VR$. There is no reflected classical trajectory, but
only one corresponding to the transmitted particle.
 The \textit{Zitterbewgung} of the quantum mean trajectory is still visible in the intermediary region $(\yL,\yR)$,
 but it clearly diminishes for higher and higher energies above the top potential value $\VR$,
together with an improvement of
the coincidence of the quantum trajectory with the classical one.

%%%%%%%%%%%%%%%%%%%%%%%%%%%%%%%%%%
\subsubsection{Approximatively constant electric field}\label{Const field}

\begin{figure}[htb]
\centering
\includegraphics[scale=.56]{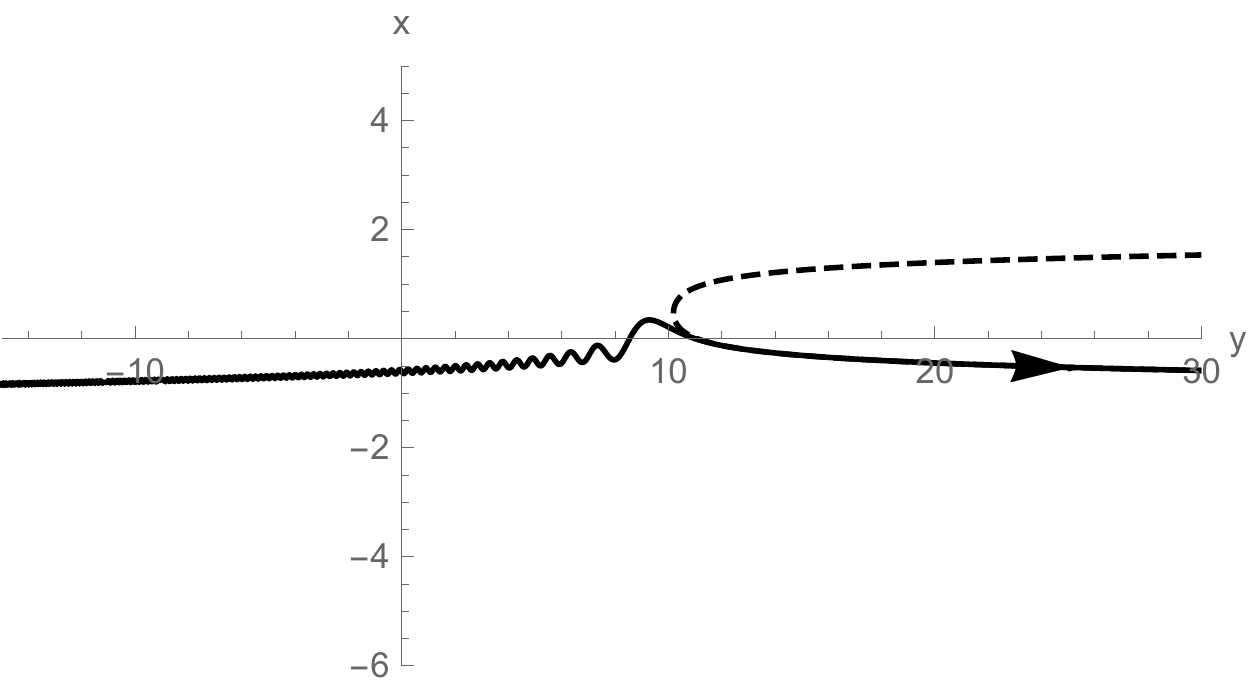}\hspace{4mm}
\includegraphics[scale=.56]{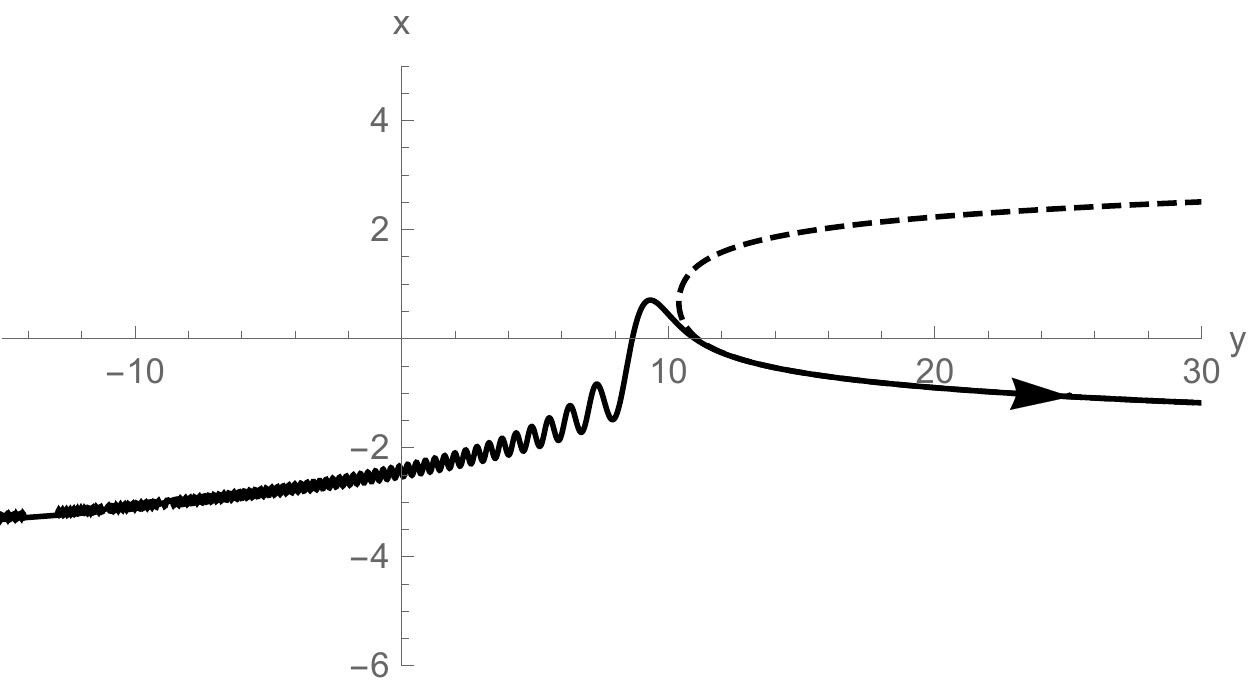}\hspace{4mm}
\caption{\small 
Step potential as specified in the text, with a large scale parameter 
$L=900$. Quantum mean trajectory
for a CB particle coming  from the left and emerging after the 
potential barrier as a VB particle is shown. The dashed line shows 
the classical trajectory 
of a  particle of negative kinetic energy coming from the right and 
repelled by the same electric field. $(\om,k)$ = $(10.0,\,0.2)$ 
and $(10.0,\,0.4)$.
}
\label{fig_const-E}
\end{figure}

The interaction of the particle of charge $q$ with a
constant electric field $E$ in the $y$ direction would be given 
by the potential
\eq
V(\bx)=-q Ey.
\eqn{constEpot}
Nevertheless, in order to take advantage of the calculation apparatus used 
in the preceding subsection, we simulate the situation with an 
oblique step potential whose domain of non-triviality extends 
to large positive and negative values of the $y$ coordinate. 
More specifically, we choose the following expressions for the 
potential parameters defined in Fig. \ref{potgeneric}: 
$\{\yL,\,\yL',\,\yR',\,\yR\}$
= $\{-L,\,L,\,L,\,L\}$ and $\{\VL,\,V_0,\,\VR\}$
= $\{-L,\,L,\,L\}$, where the scale $L$ is ``large''.
This means that,  the charge of the particle being $q=1$, we have 
a constant electric field $E=-1$ in the 
interval $-L<y<L$, and $E=0$ outside of this interval.
 Thus in a region which is 
reasonably small with respect to the scale $L$ and located 
far from the boarder $\{-L,L\}$, as in Fig. \ref{fig_const-E}, 
where $L$ has been given the value 900,
the behaviour of the particle must approximate the behaviour 
it would have for a really constant field. Moreover, in order 
to take into account the part of the trajectory where the 
quantum behaviour differs significantly from the classical one, 
we must take  values for $|\om/E|$ small with respect to the scale $L$.
The coincidence of the quantum mean trajectory with the classical one is
very good in the $y>|\om/E|$ region, whereas no such comparison 
is possible in the left region because of the superposition 
of incoming and reflecting modes  -- which is the cause of the 
observed \textit{Zitterbewegung}.
%%%%%%%%%%%%%%%%%%%%%%%%%%%%%%%%%%%%%%%%%%%%%%%%%%%%%%%%%%%%
%%%%%%%%%%%%%%%%%%%%%%%%%%%%%%%%%%%%%%%%%%%%%%%%%%%%%%%%%%%%
The reflection  probability $\RR$ is shown
in Table~\ref{table-k-R} for various values of $k_x$ and one of $\om$. 
We have checked that the results are in fact practically independent of  
the energy $\om$ if the order of magnitude of the latter is kept 
small with respect to the scale $L$.
 (It would be rigorously independent of $\om$ in the case of a 
 truly constant field as given by the potential \equ{constEpot}). 
 We also note that $\RR$ grows with $k_x$. All of this is in 
 qualitative accord with the
\textit{plateaux} in $\om$ observed in the three examples shown 
in Fig. \ref{fig_obl-step_R&T}, as well as with the $k_x$ dependence 
of these \textit{plateaux}. 
%%%%%%%%%%%%%%%%%%%%%%%%%%%%%%%%%%
\begin{table}[htb]
\vspace{5mm}

\begin{center}\begin{tabular}{|l|lllll|}%{|l||l|l|l|l|l|}
\hline
$k_x$&0.0&0.1&0.2&0.4&0.8\\
%\hline
$\RR$&0.000&0.031&0.118&0.395&0.866\\
\hline
\end{tabular}\end{center}
\vspace{-5mm}

\caption{\small Values of the reflection coefficient $\RR$ for $\om=10$ and various 
values of $k_x$.}
\label{table-k-R}\end{table}
%%%%%%%%%%%%%%%%%%%%%%%%%%%%%%%%%%

%%%%%%%%%%%%%%%%%%%%%%%%%%%%%%%%%%
\subsubsection{Oblique barrier potential}\label{barrier potential}

\begin{figure}[htb]
\centering
\includegraphics[scale=0.55]{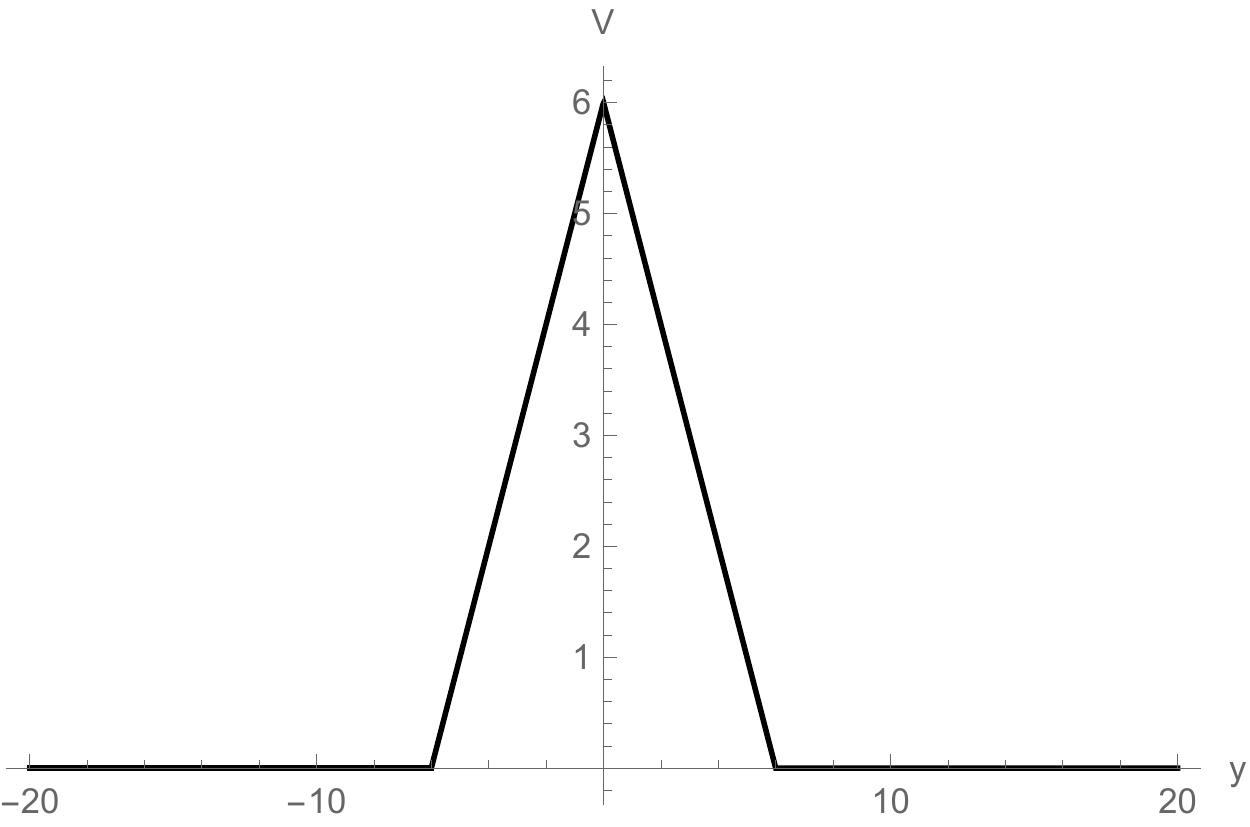}
%\hspace{5mm}
\caption{\small 
 \hfill Oblique barrier potential. The values of the parameters of 
 \equ{generic-pot} are taken as
 $\yL$ = $-6$, $\yL'$ = $\yR'=0$, $\yR=6$, $\VL=\VR=0$, $V_0=6$.
}
\label{figbarrier}
\end{figure}

\begin{figure}[htb]
%Mathematic  file "thin2 obl barr - R, T  LISTA for paper.nb"
\centering
\includegraphics[scale=0.54]{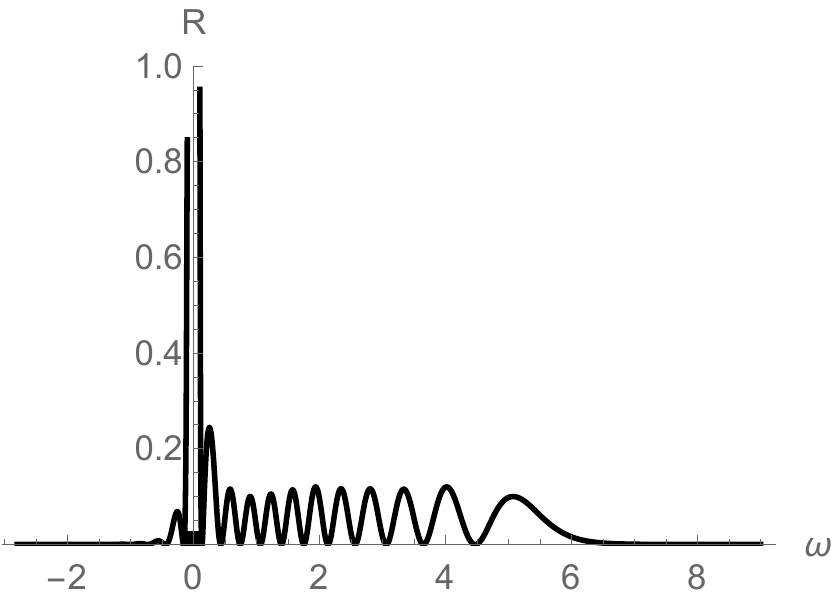}\hspace{4mm}
\includegraphics[scale=0.54]{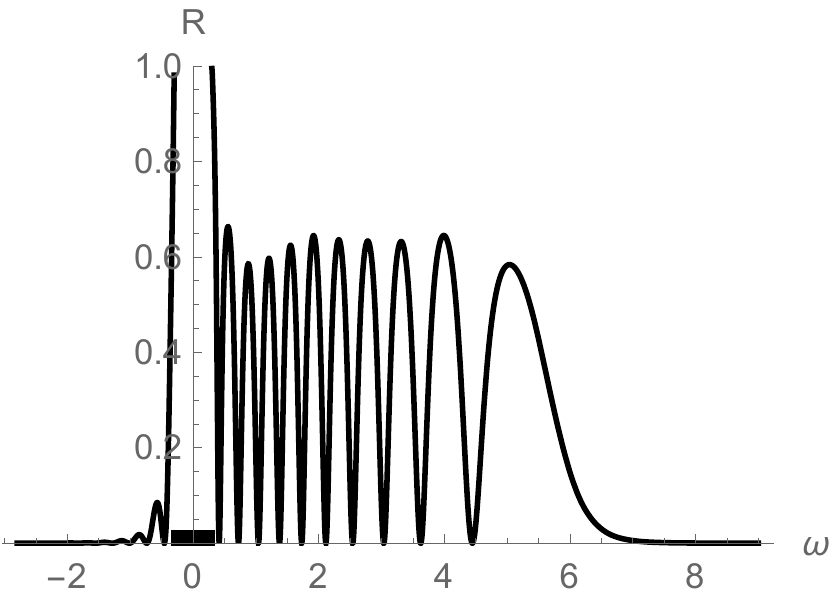}\hspace{4mm}
\includegraphics[scale=0.54]{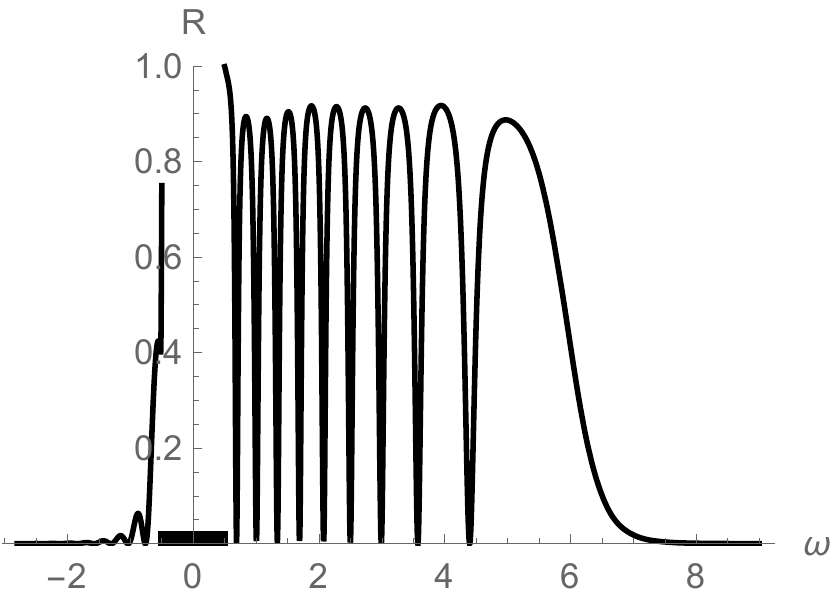}%\hspace{4mm}
\caption{\small 
 Potential of Fig. \ref{figbarrier}:  Reflection probability $\RR$
 as a function of the frequency $\om$ in the massless case for $k_x=0.1$, 
 $0.3$ and  $0.5$, respectively. 
 The heavy horizontal segment shows the 
 energy gap interval.  
}
\label{fig_obl-barr_R&T}
\end{figure}

We consider now a barrier potential $V(y)$, a smoothed version of the 
one shown in Fig. \ref{figbarrier}. 
%%%%%%%%%%%%%%%%%%%%%%%%%%%%%%%%%%%%%%%%%%%%%%%
%%%%%%%%%%%%%%%%%%%%%%%%%%%%%%%%%%%%%%%%%%%%%%%
Fig. \ref{fig_obl-barr_R&T} shows the reflection probability $\RR$ as a
function of the energy $\om$ for various values of the $x$-component $k_x$
of the momentum.
%%%%%%%%%%%%%%%%%%%%%%%%%%%%%%%%%%%%%%%%%%%%%%%
%%%%%%%%%%%%%%%%%%%%%%%%%%%%%%%%%%%%%%%%%%%%%%%
The transmission resonance oscillations of the reflection coefficient 
$\RR$ seen in the case of the square potential  barrier 
(see Fig. \ref{fig_sq-barrier}) appear here, too. $\RR$ oscillates between 0 
(full transparency) to a maximum value which depends on the energy $\om$ 
and tends to decrease together with the value of the 
momentum $x-$ component $k_x$, going to 0 in the limit $k_x=0$.

\begin{figure}[htb]
%Mathematica file "thin2 obl barr - traj_om=4 a 7 for paper.nb"
\centering
\includegraphics[scale=0.5]{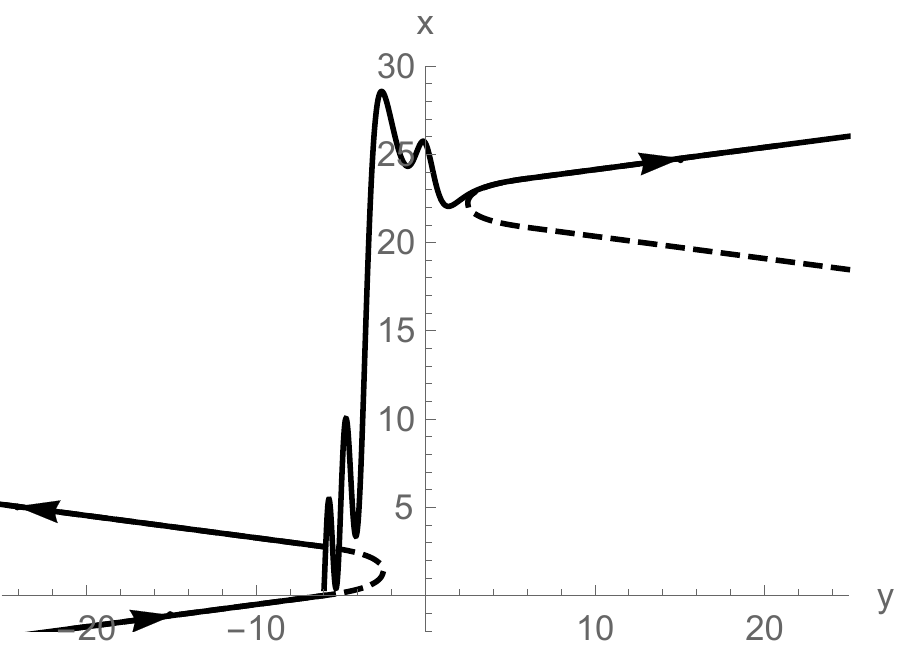}\hspace{4mm}
\includegraphics[scale=0.5]{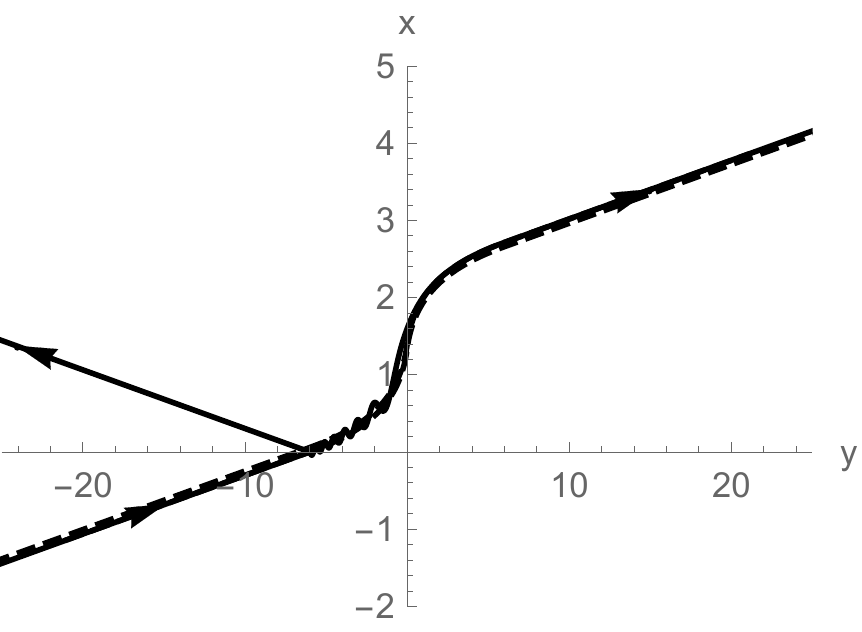}\hspace{4mm}
\includegraphics[scale=0.5]{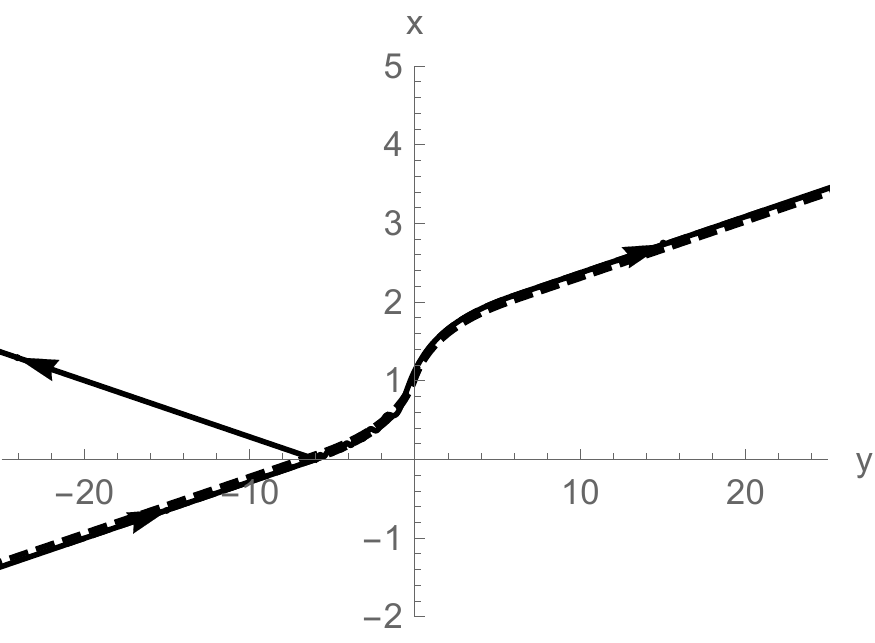}%\hspace{4mm}
\caption{\small 
 Potential of Fig. \ref{figbarrier}: Quantum mean trajectories 
 (continuous lines)
 shown together with the corresponding classical trajectories 
 (dashed lines), for $(\om,k_x)$ = $(4.0,0.5)$, $(6.6,0.5)$ and $(7.0,0.5)$.
 The corresponding values of the reflection coefficient $\RR$ are
 0.913, 0.072 and 0.018, respectively.
}
\label{fig_obl-barr_traj}
\end{figure}

Fig. \ref{fig_obl-barr_traj} shows the quantum mean trajectories 
compared with the corresponding classical ones for one value of $k_x$ and 
three values of the energy $\om$. 
For the first case, with a very small transmission probability, 
$\TT=0.087$,
we show the classical trajectory of a incident particle from 
the left and  reflected by a negative electric field, as well as that of an 
particle incoming from the right and 
reflected by a positive electric field. For the other two cases, where 
there is no reflection at the classical level, we show the trajectory 
of the classical particle going through.

We observe a very good coincidence of the classical and mean 
quantum trajectories, with the exception, in the first case, 
of a small part of the interaction region where quantum effects 
are preponderant. 
%%%%%%%%%%%%%%%%%%%%%%%%%%%%%%%%%%%%%%%%%%%%%%%%%%%%%%%%%%%%
%%%%%%%%%%%%%%%%%%%%%%%%%%%%%%%%%%%%%%%%%%%%%%%%%%%%%%%%%%%%

 The classical trajectories in  the first case exhibit the 
classical Klein phenomenon mentioned 
at the end of Appendix \ref{app-electrostatic field}: although the particle cannot go 
through the barrier, it may either come from the left and be repulsed to the left, 
having a positive kinetic energy, or  it may either come from the right and be 
repulsed to the right having  a negative kinetic energy.

\begin{figure}[htb]
\centering
\includegraphics[scale=0.7]{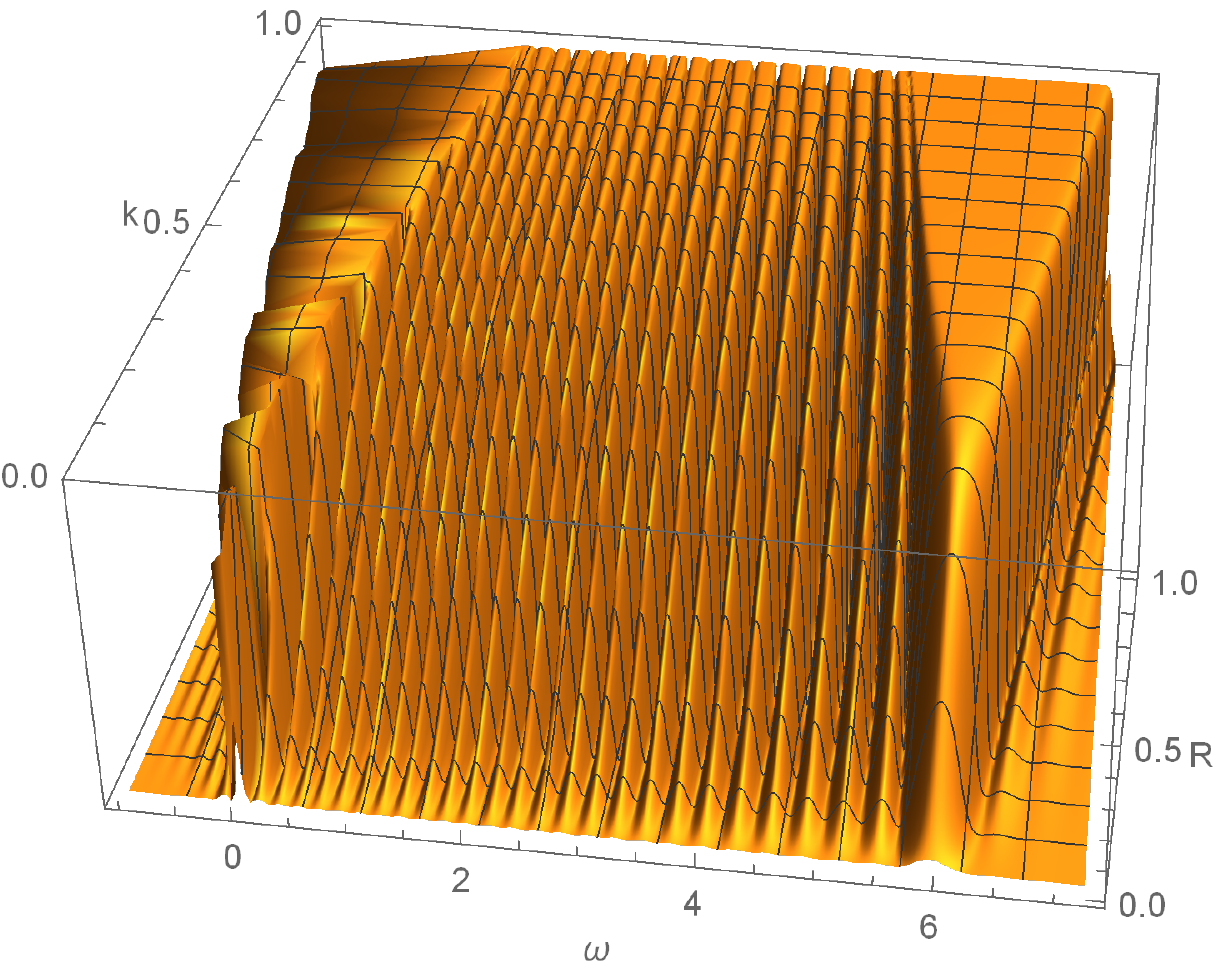}\hspace{4mm}
\caption{\small 
Reflection probability $\RR$ in function of $\om$ and $k_x$ 
for the potential barrier parameters 
 $\yL=-10$, $\yL'=-4$, $\yR'=4$, $\yR=10$, $\VL=\VR=0$, $V_0=6$
 (see Eq. \equ{generic-pot} and Fig. \ref{potgeneric}).
}
\label{fig3D_barrier_1}
\end{figure}

Fig. \ref{fig3D_barrier_1} shows the reflection probability $\RR$
as a function of both the energy $\om$ and the momentum component $k_x$,
for another choice of the potential parameters.

%%%%%%%%%%%%%%%%%%%%%%%%%%%%%%%%%%%%q%%%%%%%%%%%%%%%%%%%%%%%%

%%%%%%%%%%%%%%%%%%%%%%%%%%%%%%%%%%
\subsubsection{Oblique well potential}\label{well potential}

\begin{figure}[htb]
\centering
\includegraphics[scale=0.55]{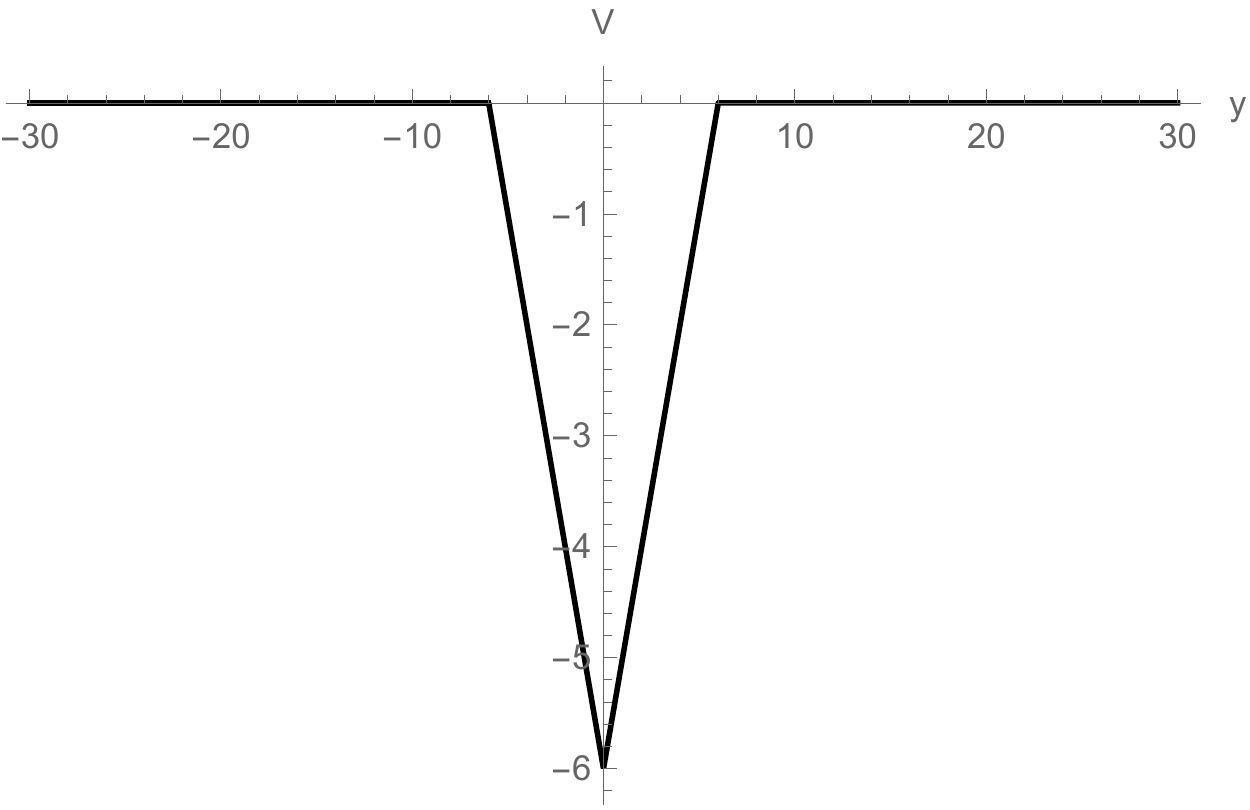}
%\hspace{5mm}
\caption{\small 
 \hfill Oblique well potential. The values of the parameters of \equ{generic-pot} are taken as
 $\yL=-6$, $\yL'=\yR'=0$, $\yR=6$, $\VL=\VR=0$, $V_0=-6$.
}
\label{figwell}
\end{figure}

\begin{figure}[htb]
%Mathematica file "thin2 obl barr - traj_om=4 a 7 for paper.nb"
\centering
\includegraphics[scale=0.38]{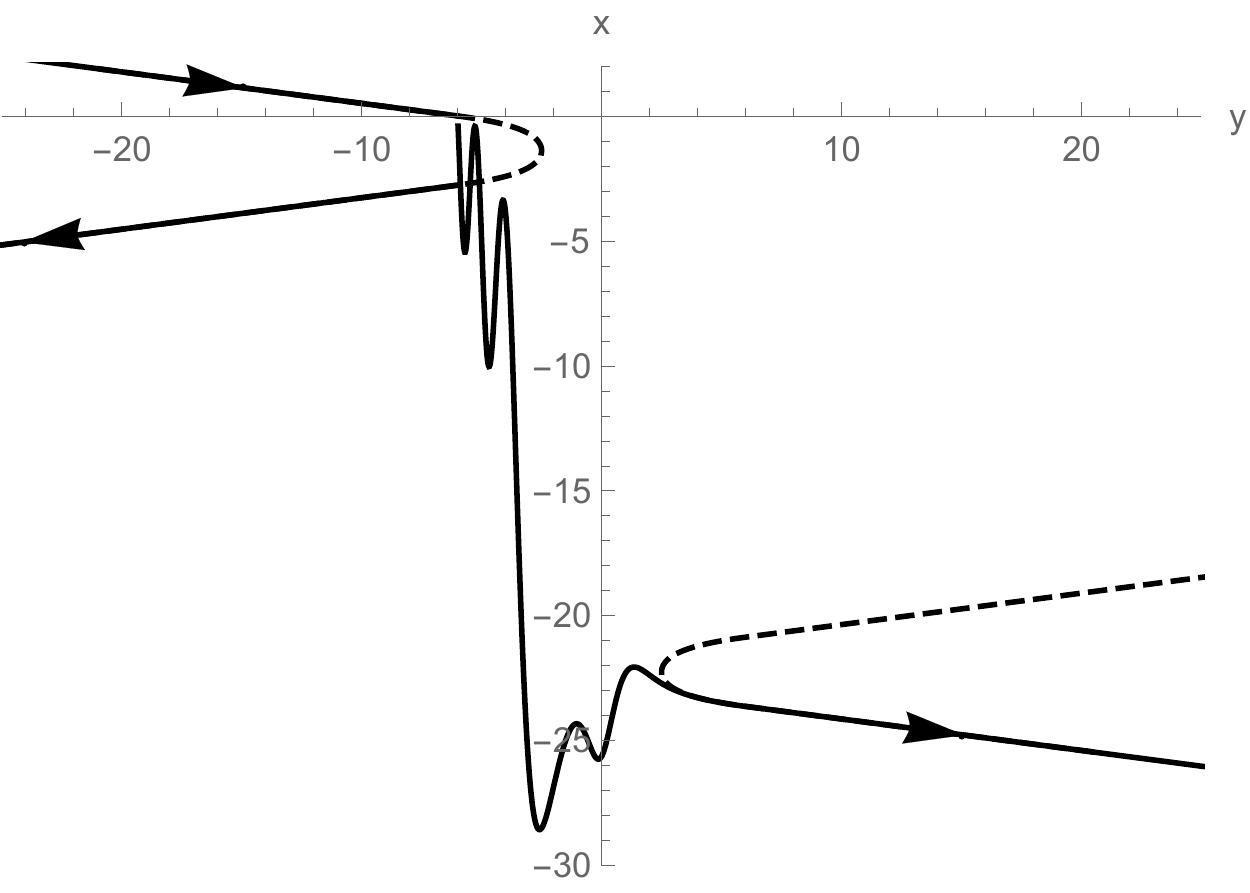}\hspace{4mm}
\includegraphics[scale=0.38]{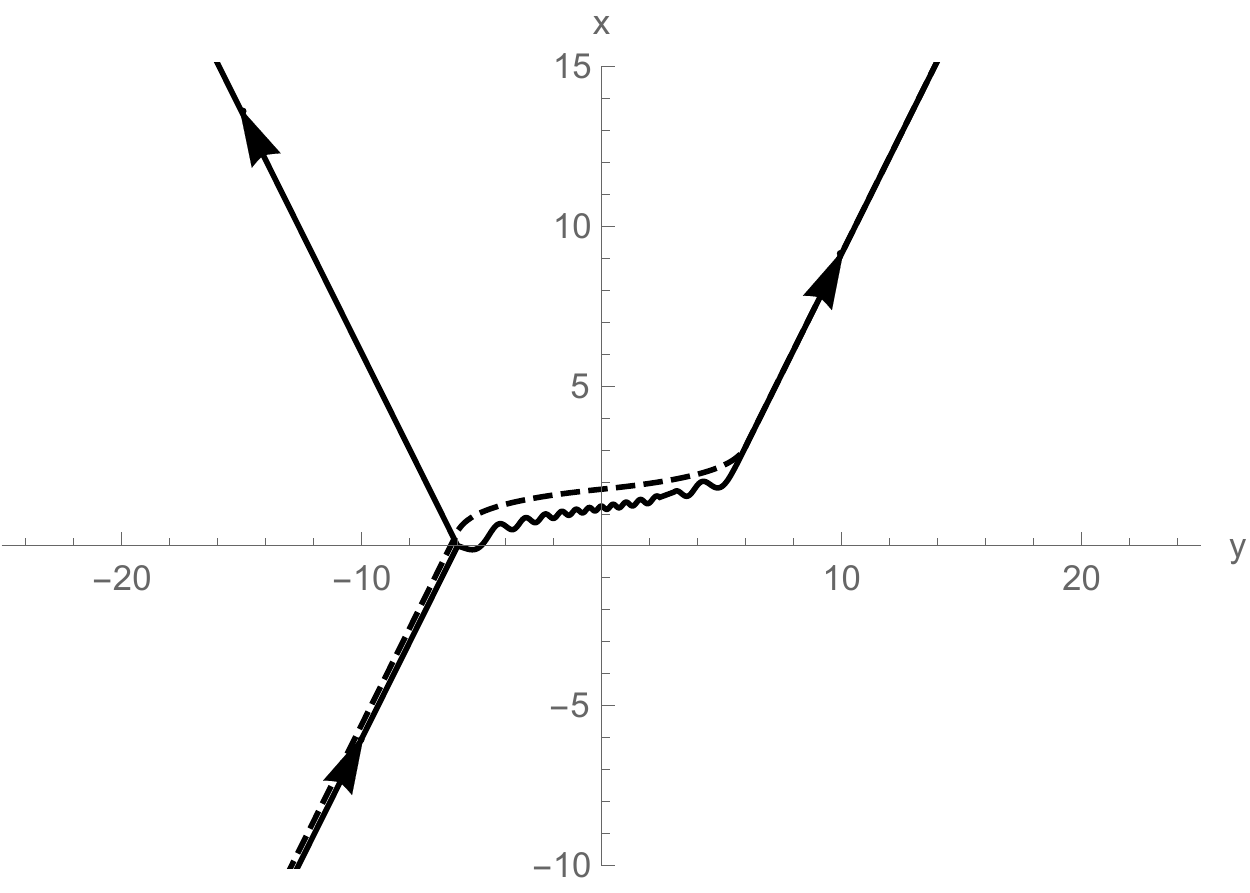}%\hspace{4mm}
\includegraphics[scale=0.38]{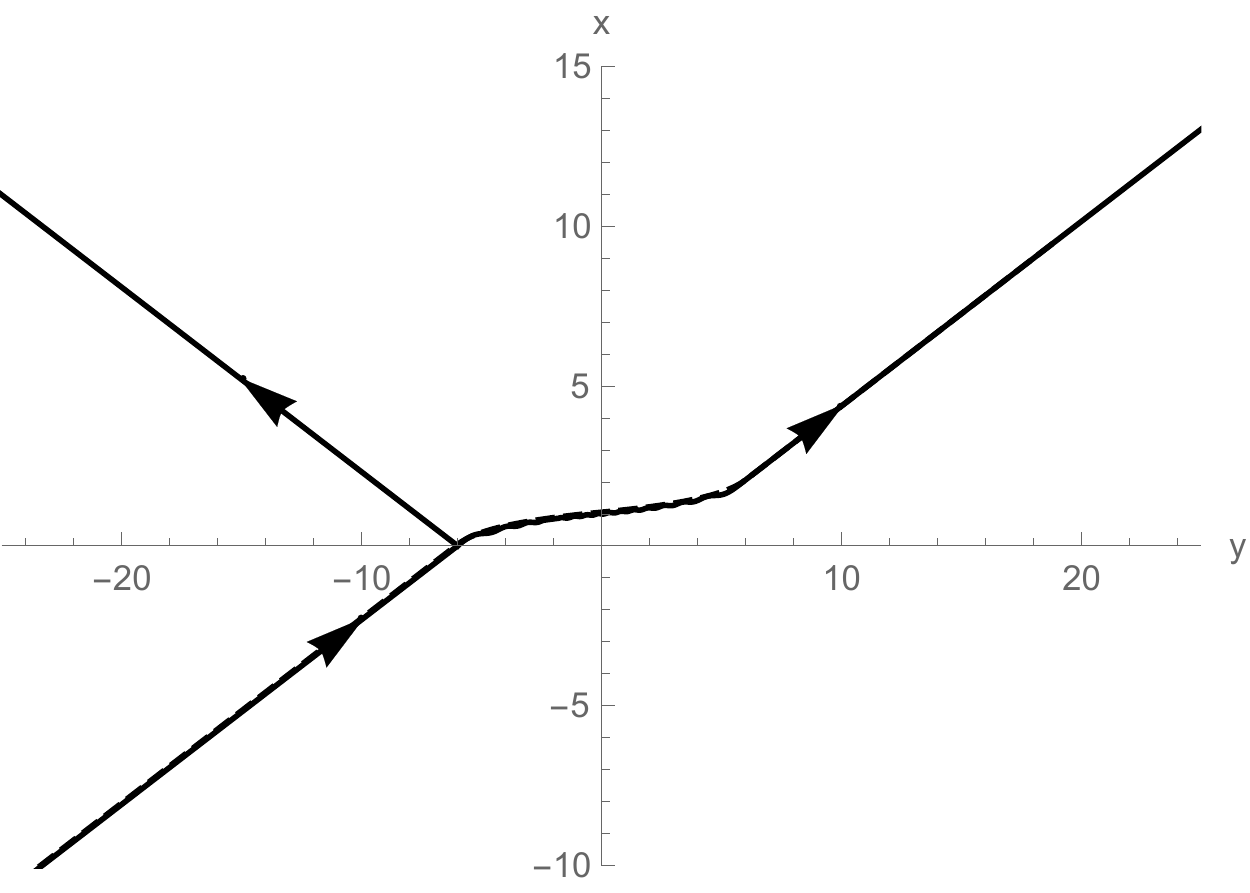}%\hspace{4mm}
\caption{\small 
 Potential of Fig. \ref{figwell}: Quantum mean trajectories 
 (continuous lines)
 shown together with the corresponding classical trajectories 
 (dashed lines), for $(\om,k_x)$ = $(-4.0,0.5)$, $(0.6,0.5)$ and $(1.0,0.5)$.
 The corresponding values of the reflection coefficient $\RR$ are
 0.913, 0.385 and 0.006, respectively.
}
\label{fig_obl-well_traj}
\end{figure}

The case of the well potential depicted in Fig. \ref{figwell} 
%%%%%%%%%%%%%%%%%%%%%%%%%%%%%%%%%%%%%%%%%%%%%%%
%%%%%%%%%%%%%%%%%%%%%%%%%%%%%%%%%%%%%%%%%%%%%%%
is symmetric
to that of the barrier potential of Subsection \ref{barrier potential} 
due to the invariance of the theory under charge conjugation. 
This means, for the chosen parametrizations of both potentials, that 
 a CB (or VB) particle of energy $\om$ 
submitted to the barrier potential  and VB (or CB) particle of 
energy $-\om$ submitted to the well potential, both with the same value 
of the $k_x$ component, will have a symmetric behaviour. 
In particular they will have equal reflection and transmission 
probabilities and follow symmetric mean quantum trajectories. The latter is exemplified 
by the comparison of the first graph of Fig. \ref{fig_obl-barr_traj} 
with the first graph of Fig. \ref{fig_obl-well_traj}
%%%%%%%%%%%%%%%%%%%%%%%%%%%%%%%%%%%%%%%%%%%%%%%%%%%%%%%%%barr%%%
%%%%%%%%%%%%%%%%%%%%%%%%%%%%%%%%%%%%%%%%%%%%%%%%%%%%%%%%%%%%
The second and third graphs of the latter figure show a particle flying over the well.

%%%%%%%%%%%%%%%%%%%%%%%%%%%%%%%%%%%%%%%%%%%%%%%%%%%%%%%%%%%%

%%%%%%%%%%%%%%%%%%%%%%%%%%%%%%%%%%
%%%%%%%%%%%%%%%%%%%%%%%%%%%%%%%%%%
%%%%%%%%%%%%%%%%%%%%%%%%%%%%%%%%%%
%%%%%%%%%%%%%%%%%%%%%%%%%%%%%%%%%%

%%%%%%%%%%%%%%%%%%%%%%%%%%%%%%%%%%

%%%%%%%%%%%%%%%%%%%%%%%%%%%%%%%%%%
\section{Conclusions}

We have examined various examples of a relativistic quantum 
massless spinning particle in two-dimensional space, 
submitted to an electrostatic field oriented in one direction 
-- the $y$-coordinate direction. These examples are characterized 
by $y$-dependent potentials
 of the form of a step, a barrier or a well. 
In each case we have computed the stationary solutions 
of the corresponding Dirac equation, together with the 
reflection and transmission coefficients. We have also 
computed in  most cases the quantum mean trajectories and 
compared them with their classical counterparts, obtained 
by integration of the classical equations of motion, with 
boundary conditions  adjusted to the quantum solution.

The explicit solutions found in the 
literature~\cite{stander:2009, Dragoman_2008, Katsnelson_2006, Calogeracos:1999yp} 
 concern a particle 
 %moving in a one-dimension space and being 
 submitted to a square potential. Those of them which avoid 
 the Klein ``paradox'' problem by properly taking into account 
 the characteristics of the object being a CB or VB particle, 
  \ie a particle of positive or negative kinetic energy,
 turn out to coincide with ours. 
 Examining the momentum and energy dependence of 
 the reflection and transmission coefficients of our solutions 
 for more general potentials such as smoothed oblique steps, 
 barriers and wells, we found a behaviour of these coefficients 
 which is  qualitatively similar to that of the square potentials. 
 In particular we reproduce explicitly in each case the 
 Klein phenomenon of transmission at values of the energy 
 for which the non-relativistic particle wave function 
 would be exponentially damped through the barrier or behind the step.

Concerning the comparison of the quantum mean trajectories 
with the classical one, we found a very good agreement, 
excepted in situations where a non-negligible 
\textit{Zitterbewegung} is present due to interference 
between right and left moving modes.

 One important commentary on the Klein 
phenomenon which we 
observe in our calculations is still deserved, as, \eg 
in the case of the potential of Fig. \ref{figstep}. 
In a non-relativistic theory, if the energy is below 
the top of the potential, there is no possibility of 
the particle to move in the right region, neither 
classically, nor quantically - excepted for an evanescent
 wave function in this region in the quantum case. As our 
 calculations confirm, in the same setting, the transmission 
 probability may be large in the relativistic case. It is of 
 course zero in the classical relativistic theory, but there 
 are solutions for the particle moving in the right region 
 (see the first graphic of Fig. \ref{fig_obl-step_traj}), with an acceleration 
 opposed to the electric force due to a negative  kinetic 
 energy, which plays the role of an inertial factor. 
 This is what we could call a ``classical Klein phenomenon''.

%%%%%%%%%%%%%%%%%%%%%%%%%%%%%%%%%%%%%%%%%%
\subsubsection*{Acknowledgements}
This work was partially funded by the
 Conselho Nacional de Desenvolvimento Cien\-t\'{\i}\-fi\-co e
 Tecnol\'{o}gico---CNPq, Brazil (I.M., B. N., Z.O.  and O.P.),
by the Coordena\cao\ de Aperfei\c coamento de Pessoal de N\ii vel 
Superior---CAPES, Brazil (I.M. and B.N.),
by the Fundação de Amparo a Pesquisa do Estado de 
Minas Gerais - FAPEMIG, Brazil (O.P.)
and by the Grupo de Sistemas Complejos de la Carrera de Física 
de la {Universidad Mayor de San Andr\'es, UMSA, Bolivia (Z.O.)}, for their 
support.

%%%%%%%%%%%%%%%%%%%%%%%%%%%%%%%%%%%%%%%%%%

%%%%%%%%%%%%%%%%%%%%%%%%%%%%%%%%%%%%%%%%%%%%
\appendix
\section*{Appendices}

%%%%%%%%%%%%%%%%%%%%%%%%%%%%%%%%%%%%%%%%%%%%
\section{Notations and conventions}\label{not-conv}

\eq\ba{ll}
\mbox{Units are such that} &c = \hbar = 1,\es
\mbox{Space-time coordinates:}  &(x^\m,\,\m=0,1,2),\quad
         \bx=(x^a,\,a= 1,2),\es
\mbox{Space-time metric:} &\eta_{\m\n} =\mbox{diag}(1,-1,-1),\es
\mbox{Dirac matrices:} &
\g^0=\s_z,\, \g^1=i\s_x,\,\g^2=i\s_y\\
&\mbox{os $\s$'s are the Pauli matrices)},\es
\a\mbox{ - matrices:}  & \a^i=\g^0\g^i,\, a=1,2,\es
\mbox{Conjugate spinor:} & \bar{\psi}=\psi^\dagger\gamma^0.
\ea\eqn{conventions}

%%%%%%%%%%%%%%%%%%%%%%%%%%%%%%%%%%%%%%%%%%%%%%%%%%
\section{Classical equations in the case of a
 \texorpdfstring{$y$}{Lg}-dependent electrostatic 
 field}\label{app-electrostatic field}

We consider here a massless particle of charge $q=1$ in the presence of an
 electrostatic field ${\bf E}=(0,-V'(y))$ derived from the 3-potential 
$A=(V(y),0,0)$ which depends only on the space coordinate $y$.
The equations
are given by \equ{eom}, with $m=\p_5=0$ and the partial gauge fixing 
$\chi=0$.  We restrict ourselves  
on solutions with the spin variables $\p^\m=0$. With the choice of 
the worldline parametrization\footnote{We use the notation $x^0=t,\ x^1=x,\ x^2=y$
for the space-time coordinates.}  
$\la=t\ (t=x^0)$, the second of Eqs.
\equ{eom} then yields the constraint
 \eq
 \dot{x}^2+\dot{y}^2 = 1,
 \eqn{class-constr}
\ie the velocity is that of light.
The first of Eqs \equ{eom} for $\m=0$ yields the conservation of the total 
energy $\om$: $\dot\om =0$, where
\[
\om=\dfrac{1}{e(t)}+V(y(t)).
\]
For $\m=1,2$, we get
\eq\begin{array}{l}
(\om - V(y)) \dot{x} - (\om -V({\bar y})) \bar{\dot{x}} = 0,\es
(\om - V(y)) \ddot{y} 
+ (1 - (\dot{y})^2) V'(y)=0.
\end{array}\eqn{class-eq-E}
The first of these equations has been obtained by integrating the 
corresponding second order equation thanks to energy conservation and to the 
$x$-independence of the potential, with 
 ${\bar y}=y({\bar t})$ and $\bar{\dot{x}}=\dot{x}({\bar t})$ 
as initial values at some initial time ${\bar t}$.

Solutions of the equations of motion \equ{class-eq-E} are uniquely 
determined by giving 
3 boundary conditions, which may be the values of ${\bar y}$, $\bar{\dot{x}}$ and  ${\bar x}=x({\bar t})$, assuming the validity of \equ{class-constr} at ${\bar t}$.

 It is worthwhile to note that, in the second equation \equ{class-eq-E},
the kinetic energy factor $\om_{\rm kin}(y)$ = $\om-V(y))$, which can be positive or negative
depending on the position $y$, plays the role of an inertia 
coefficient~\cite{Morales:2017rlk}. In particular, the sign 
of the $y$-component 
of the acceleration will depend on the sign of $\om_{\rm kin}(y)$.
 We may be consider this as a ``classical Klein phenomenon''.

%%%%%%%%%%%%%%%%%%%%%%%%%%%%%%%%%%%%%%%%%%

\end{document}